# Network Global Testing by Counting Graphlets

Jiashun Jin [1]   Zheng Tracy Ke [2]   Shengming Luo [1]

## Abstract

Consider a large social network with possibly severe degree heterogeneity and mixed-memberships. We are interested in testing whether the network has only one community or there are more than one communities. The problem is known to be non-trivial, partially due to the presence of severe degree heterogeneity. We construct a class of test statistics using the numbers of short paths and short cycles, and the key to our approach is a general framework for canceling the effects of degree heterogeneity. The tests compare favorably with existing methods. We support our methods with careful analysis and numerical study with simulated data and a real data example.

## 1. Introduction

Given a large symmetrical network, we are interested in the global testing problem where we use the adjacency matrix of the network to test whether the network consists of only one community or that it consists of multiple communities, where some nodes may have mixed memberships.

Real networks frequently have *severe degree heterogeneity*. The Stochastic Block Model (SBM) is well-known, but does not accommodate severe degree heterogeneity. To tackle the problem, Karrer and Newman (2011) proposed the Degree-Corrected Block Model (DCBM). DCBM strikes a better balance between theory and practice than SBM, and has become increasingly more popular recently.

We adopt a *Degree-Corrected Mixed-Membership (DCMM)* model (Jin et al., 2017). DCMM can be viewed as an extension of DCBM, but allows for mixed memberships. Suppose the network has $n$ nodes and $K$ perceivable communities

$$\mathcal{C}_1, \mathcal{C}_2, \ldots, \mathcal{C}_K.$$

[*]Equal contribution   [1]Department of Statistics and Data Science, Carnegie Mellon University, Pittsburgh, USA   [2]Department of Statistics, University of Chicago, Chicago, USA. Correspondence to: Jiashun Jin <jiashun@stat.cmu.edu>.

For each node, we assign a Probability Mass Function (PMF) $\pi_i = (\pi_i(1), \pi_i(2), \cdots, \pi_i(K))'$, where $\pi_i(k)$ is the "weight" that node $i$ puts on community $\mathcal{C}_k$, $1 \leq k \leq K$. We call node $i$ "pure" if $\pi_i$ is degenerate and "mixed" otherwise. Let $A \in \mathbb{R}^{n,n}$ be the adjacency matrix, where $A_{ij} = 1$ if nodes $i$ and $j$ have an edge, and $A_{ij} = 0$ otherwise (all diagonal entries of $A$ are 0 as we don't treat a node as connecting to itself). In DCMM, we think the upper triangle of $A$ contains independent Bernoulli random variables. Moreover, for $n$ degree heterogeneity parameters $\theta_1, \theta_2, \ldots, \theta_n$ and a non-singular, irreducible matrix $P \in \mathbb{R}^{K,K}$,

$$\mathbb{P}(A_{ij} = 1) = \theta_i \theta_j \sum_{k=1}^{K} \sum_{\ell=1}^{K} \pi_i(k) \pi_j(\ell) P_{k\ell}. \quad (1)$$

Also, we assume (a) all diagonals of $P$ are 1, and (b) each of the $K$ communities has at least one pure node. With such constraints, DCMM is identifiable (Jin et al., 2017).

Let $\Theta$ be the $n \times n$ diagonal matrix $\Theta = \mathrm{diag}(\theta_1, \ldots, \theta_n)$ and let $\Pi$ be the $n \times K$ matrix $\Pi = [\pi_1, \pi_2, \ldots, \pi_n]'$. Let $\Omega = \Theta \Pi P \Pi' \Theta$ (recall that $P \in \mathbb{R}^{K,K}$):

$$\Omega = \begin{bmatrix} \theta_1 & & \\ & \ddots & \\ & & \theta_n \end{bmatrix} \begin{bmatrix} \pi_1' \\ \vdots \\ \pi_n' \end{bmatrix} P \, [\pi_1, \ldots, \pi_n] \begin{bmatrix} \theta_1 & & \\ & \ddots & \\ & & \theta_n \end{bmatrix}.$$

Let $\mathrm{diag}(\Omega)$ be the diagonal matrix where the $i$-th diagonal entry is $\Omega_{ii}$, and let $W = A - [\Omega - \mathrm{diag}(\Omega)]$. We have

$$A = [\Omega - \mathrm{diag}(\Omega)] + W = \text{``signal''} + \text{``noise''}.$$

**Remark 1**. Many recent works use the "Random Degree Parameter (RDP)" model (which is narrower than ours): fixing a scaling parameter $\alpha_n > 0$ and a density function $f$ over $(0, \infty)$ where the first a few moments of $f$ are finite, and especially the second moment is 1, we assume $(\theta_i/\alpha_n) \stackrel{iid}{\sim} f$, $i = 1, 2, \ldots, n$. We call the resultant DCMM model the DCMM-RDP model. In applications where we don't know how $\theta_i$'s are correlated, it is preferable to treat $\theta$ as non-random as before, and it is safer to use the original DCMM than DCMM-RDP.

The testing problem can be cast as testing a null hypothesis $H_0^{(n)}$ against a specific hypothesis $H_1^{(n)}$ in its complement:

$$H_0^{(n)}: K = 1 \qquad vs. \qquad H_1^{(n)}: K > 1, \quad (2)$$

where under $H_1^{(n)}$, DCMM holds for some eligible $(P, \theta_1, ..., \theta_n, \pi_1, ..., \pi_n)$. Note that under $H_0^{(n)}$, $P = 1$, $\pi_1, ..., \pi_n$ are all degenerate, and $\mathbb{P}(A(i,j) = 1) = \theta_i \theta_j$.

Most existing literatures for the testing problem (2) have been focused on the special case where

- *No degree heterogeneity.* $\theta_1 = \theta_2 = \ldots = \theta_n$.
- *No mixed-membership.* All $\pi_i$ are degenerate PMFs.

Many tests were proposed for this special case, including but are not limited to the likelihood ratio test approach (Wang & Bickel, 2017) and the spectral approach (Bickel & Sarkar, 2016; Lei, 2016; Banerjee & Ma, 2017).

However, in our setting, $\theta_1, \theta_2, \ldots, \theta_n$ vary significantly from one to another, and it is unclear how to extend the methods above to the current setting. The likelihood ratio test is not applicable, for there are many unknown parameters $(\theta_1, \pi_1), (\theta_2, \pi_2), ..., (\theta_n, \pi_n)$. The spectral approach also faces challenges, because the distributions of such test statistics depend on unknown parameters in a complicated way, and it is nontrivial to figure out the rejection region.

Also, it may be tempting to adapt the recent approaches on estimating $K$ to our testing problem (Saldana et al., 2017; Chen & Lei, 2017; Le & Levina, 2015), but similarly, due to severe degree heterogeneity, the null distributions of such statistics are not tractable, so they cannot be used directly.

Recently, Gao and Lafferty (2017) (see also Bubeck et al. (2016) which is related but on different settings) proposed a new test called the Erdős-Zuckerberg (EZ) test for DCMM, with the following constraints.

- (GL1) No mixed-membership: all $\pi_i$ are degenerate.
- (GL2) The community labels are uniformly drawn so the $K$ communities have roughly equal sizes.
- (GL3) The DCMM-RDP holds (i.e., $\theta_i$ are iid samples), and the $K \times K$ matrix $P$ has the special form of

$$P = \begin{bmatrix} a & b & \cdots & b \\ \vdots & & \ddots & \vdots \\ b & b & \cdots & a \end{bmatrix}.$$

Note that the last bullet point is particularly restrictive.

Gao and Lafferty (2017) made an interesting observation that the effect of the degree heterogeneity parameters $\theta_1, \theta_2, \ldots, \theta_n$ is largely canceled out in the EZ test, and the test statistic approximately equals to 0 under the null, regardless of what $\theta_1, \theta_2, \ldots, \theta_n$ are. This allows us to find a convenient way to map out the rejection region.

Unfortunately, the authors did not make it clear whether the cancellation is "coincidental" and is due to the symmetry they imposed on the model (see GL1-GL3), or is "inherent" and holds for much broader settings.

In this paper, we introduce a class of test statistics by counting the number of graphlets in the network. Fixing a small $m \geq 1$, we count two kinds of graphlets: length-$m$ paths and $m$-cycles. Our main contributions are as follows:

- *Ideation.* For a $K$-vector $\eta$ and a $K \times K$ matrix $G^{1/2}PG^{1/2}$ to be introduced, we derive succinct proxies for the number of length-$m$ paths and $m$-cycles, using $\eta$ and the eigenvalues and eigenvectors of $G^{1/2}PG^{1/2}$. The proxies motivate a systematic way of constructing tests where the degree heterogeneity is largely removed so the distributions are more tractable.

  To the best of our knowledge, such proxies have not been discovered in the literature.

- *Methods and theory.* We propose a class of graphlet count (GC) test statistics, and derive their asymptotic distributions, under the null and alternative hypotheses.

  We try to be as general as possible, and our methods and theory are for DCMM with minimal constraints.

The way we construct our statistics is to use the proxies aforementioned, and thus is different from that in Gao and Lafferty (2017), which uses calculations that heavily depend on the imposed constraints GL1-GL3. See Section 3.2 for more comparison.

Our findings support the philosophy of Jin (2015) which introduced the community detection algorithm SCORE. Jin (2015) pointed out that $\theta_1, \theta_2, \ldots, \theta_n$ are required to model severe degree heterogeneity, but they turn out to be nuisance parameters, the effects of which can be largely removed with a *proper construction* of statistics.

While our tests are designed for global testing, the idea is also useful for tackling other problems. For example, we can combine our idea with those in community detection (e.g. Jin (2015), Chen et al. (2018), Qin and Rohe (2013)) to estimate the number of communities $K$.

## 2. A Class of Graphlet Count (GC) Statistics

The testing problem (2) is hard for there are so many unknown parameters: $P, \theta_1, \ldots, \theta_n, \pi_1, \ldots, \pi_n$. The parameters $\theta_1, \theta_2, \ldots, \theta_n$, which are required to model the severe degree heterogeneity of real networks, are especially hard to deal with for they vary significantly from one to another. What we need is therefore a smart test statistic that

- does not vary significantly as $\theta = (\theta_1, \theta_2, \ldots, \theta_n)'$ varies, and has a tractable limiting distribution (so it is easy to map out the rejection region),
- is powerful in differentiating the null and alternative.

Our idea is to use the graphlet-count statistics. In a network, we say a path is a "self-avoiding path" if it doesn't intersect with itself, and a "cycle" if it is a closed path that does not intersect with itself.

**Definition 2.1** For $0 \leq m \leq n$, let $B_{n,m} = \prod_{s=0}^{m-1}(n-s)$.

**Definition 2.2** For $m \geq 1$, we define the "density of length-$m$ self-avoiding paths" by

$$\widehat{L}_m = \frac{1}{B_{n,m+1}} \sum_{\substack{1 \leq i_1,\ldots,i_{m+1} \leq n \\ i_1,\cdots,i_{m+1} \text{ are distinct}}} A_{i_1 i_2} A_{i_2 i_3} \cdots A_{i_m i_{m+1}}.$$

and for $m \geq 3$, we define the "density of $m$-cycles" by

$$\widehat{C}_m = \frac{1}{B_{n,m}} \sum_{\substack{1 \leq i_1,\ldots,i_m \leq n \\ i_1,\cdots,i_m \text{ are distinct}}} A_{i_1 i_2} A_{i_2 i_3} \cdots A_{i_m i_1}.$$

We propose the family of test statistics, called the **Graphlet Count (GC)** test statistics:

$$\widehat{\chi}_{gc}^{(m)} = \widehat{C}_m - (\widehat{L}_{m-1}/\widehat{L}_{m-2})^m, \qquad m = 3, 4, \ldots. \quad (3)$$

**Remark 2**. Using the adjacency matrix $A$, $\widehat{C}_m$ and $\widehat{L}_m$ can be conveniently computed (e.g., $\widehat{C}_4 = \frac{1}{24\binom{n}{4}}[\text{tr}(A^4) - 2(1_n' A^2 1_n) + 1_n' A 1_n]$). See supplemental material.

### 2.1. The Key Idea: Why the GC Test Statistics Work

Recall that $\Omega = \Theta\Pi P \Pi'\Theta$. Let $1_n$ be the $n$-dimensional vector of 1's and let $\theta = (\theta_1, \theta_2, \ldots, \theta_n)'$. We use $(\cdot, \cdot)$ to denote the inner product of two vectors. The $K \times K$ matrix $G \equiv \Pi'\Theta^2\Pi$ and the vector $\eta \in \mathbb{R}^K$ play a key role.

**Definition 2.3** Denote the vector $G^{-1/2}\Pi'\Theta 1_n$ by $\eta$.

**Definition 2.4** For $1 \leq k \leq K$, let $\lambda_k$ be the $k$-th largest (in absolute value) eigenvalue of $G^{1/2}PG^{1/2}$, and let $\xi_k$ be the corresponding eigenvector.

It turns out that $\lambda_1, \ldots, \lambda_K$ (eigenvalues of $G^{1/2}PG^{1/2}$) are the nonzero eigenvalues of $\Omega$. The following results, which will be made precise in Theorems 3.1-3.2, play the key role:

$$n^m \cdot \widehat{C}_m \approx \text{tr}(\Omega^m) = \sum_{k=1}^K \lambda_k^m, \quad (4)$$

$$n^{m+1} \cdot \widehat{L}_m \approx 1_n' \Omega^m 1_n = \sum_{k=1}^K (\eta, \xi_k)^2 \lambda_k^m. \quad (5)$$

We explain why these equations motivate the test statistic $\widehat{\chi}_{gc}^{(m)}$. Recall that we hope to have a statistic that does not vary too much as $\theta$ varies, so first, it is desirable to remove the terms $(\eta, \xi_k)^2$, which not only depend on $\theta$ but are also not very tractable. Under the alternative hypothesis, it is unclear how to cancel these terms, but under the null hypothesis, $K = 1$, and the right hand side of (5) reduces to $(\eta, \xi_1)^2 \lambda_1^m$, and there are many ways to do the cancellation. One such way is to use the following ratio:

$$\frac{n^m \widehat{L}_{m-1}}{n^{m-1}\widehat{L}_{m-2}} \approx \frac{\sum_{k=1}^K (\eta, \xi_k)^2 \lambda_k^{m-1}}{\sum_{k=1}^K (\eta, \xi_k)^2 \lambda_k^{m-2}} \begin{cases} = \lambda_1, \text{ under } H_0^{(n)}, \\ \leq \lambda_1, \text{ under } H_1^{(n)}. \end{cases}$$
$$(6)$$

Therefore, at least under $H_0^{(n)}$, we have managed to cancel the terms $(\eta, \xi_k)^2$.

Next, it is also desirable to cancel the term $\lambda_1$, at least under the null hypothesis. Comparing (4) and (6), there are many ways to do this, and one such way is to use the $\widehat{\chi}_{gc}^{(m)}$ statistic aforementioned:

$$n^m \widehat{\chi}_{gc}^{(m)} = n^m \widehat{C}_m - [(n^m \widehat{L}_{m-1})/(n^{m-1}\widehat{L}_{m-2})]^m.$$

In fact, by (4) and (6), we have that up to a negligible term (i.e., of a smaller order than that of the standard deviation of the statistic under $H_0^{(n)}$),

$$\widehat{\chi}_{gc}^{(m)} \begin{cases} = 0, & \text{under } H_0^{(n)}, \\ \geq \frac{1}{n^m}\sum_{k=2}^K \lambda_k^m, & \text{under } H_1^{(n)}. \end{cases}$$

On the one hand, the effects of degree heterogeneity are largely canceled in the statistic, so it does not vary significantly as the vector $\theta$ varies (this is particularly important for we wish to have a rejection region that is relatively insensitive to $\theta$). On the other hand, the statistic $\widehat{\chi}_{gc}^{(m)}$ is able to differentiate the null hypothesis and the alternative hypothesis, through the term $\sum_{k=2}^K \lambda_k^m$. This suggests that $\widehat{\chi}_{gc}^{(m)}$ is a reasonable test statistic.

Note that $\widehat{\chi}_{gc}^{(m)}$ is only one of many test statistics with the desired properties above, but seemingly one of the simplest.

**Remark 3**. Recall that $\lambda_1, \ldots, \lambda_K$ are the eigenvalues of $G^{1/2}PG^{1/2}$, and they are also the eigenvalues of $PG$ (non-negative, irreducible). By Perron's theorem (Horn & Johnson, 1985), $\lambda_1$ is positive and $\lambda_1 > |\lambda_k|$ for all $2 \leq k \leq K$.

**Remark 4**. Our tests include the EZ test as a special case (i.e., $\widehat{\chi}_{gc}^{(m)}$ with $m = 3$), but our idea is by no means a straightforward extension of that in Gao and Lafferty (2017). The EZ test was derived by calculations that depend heavily on the constraints GL1-GL3 imposed on $P$, $\theta$, etc., and it was unclear whether the core idea of the EZ test is only valid when these constraints hold, or in more general settings. The GC tests are derived by (4)-(6), where the relationship between the test statistics and $\theta$, $\eta$, and the eigenvalues and eigenvectors of $G^{1/2}PG^{1/2}$ has not been discovered before, even in cases where the constraints GL1-GL3 hold.

**Remark 5**. Can we simply use $\sum_{k=2}^K \widehat{\lambda}_k^m$ as the test statistic, where $\widehat{\lambda}_k$ is the $k$-th eigenvalue of $A$? We can not, as $K$ is unknown. Can we simply use $\widehat{\lambda}_2$ as the test statistic? We can, but the asymptotic distribution of $\widehat{\lambda}_2$ is much harder to derive than that of $\widehat{\chi}_{gc}^{(m)}$ (which is Gaussian; see below), so it is challenging to determine the rejection region.

### 3. Main Results

In theory, we use $n$ as the driving asymptotic parameter, let the matrices $(\Theta, \Pi, P)$ change with $n$, and consider a sequence of problems where we test $H_0^{(n)} : K = 1$ vs. $H_1^{(n)} : K > 1$ ($K$ is unknown but does not change with $n$). Recall node $i$ is pure if $\pi_i$ is degenerate. A pure node can be in any of the $K$ communities. For $1 \leq k \leq K$, we let

$$\mathcal{N}_k = \{1 \leq i \leq n : \pi_i \text{ is degenerate and } \pi_i(k) = 1\}$$

be the set of all pure nodes in the community $k$. Assume

$$\|\theta\| \to \infty, \qquad \|\theta\|_3 \to 0. \tag{7}$$

Since $\theta_{\max} \leq \|\theta\|_3$, this implies $\theta_{\max} \to 0$. Suppose there is a constant $c_1 > 0$ so that for any $1 \leq k \leq K$,

$$\frac{\sum_{i \in \mathcal{N}_k} \theta_i^2}{\sum_{i=1}^n \theta_i^2} \geq c_1. \tag{8}$$

(this roughly says each community has sufficiently many pure nodes). Also, assume for some constant $c_2 \in (0, 1)$,

all singular values of $P$ fall between $c_2$ and $c_2^{-1}$. (9)

Denote $C_m = \mathbb{E}[\widehat{C}_m]$ and $L_m = \mathbb{E}[\widehat{L}_m]$ and introduce a non-stochastic counterpart of $\widehat{\chi}_{gc}^{(m)}$ by

$$\chi_{gc}^{(m)} = C_m - (L_{m-1}/L_{m-2})^m. \tag{10}$$

**Theorem 3.1** *Consider the DCMM model* (1) *where* (7)-(9) *hold. As* $n \to \infty$,

$$\chi_{gc}^{(m)} = \frac{1}{n^m}\left\{\sum_{k=1}^K \lambda_k^m - \left[\frac{\sum_{k=1}^K (\eta, \xi_k)^2 \lambda_k^{m-1}}{\sum_{k=1}^K (\eta, \xi_k)^2 \lambda_k^{m-2}}\right]^m\right\}$$
$$+ O(n^{-m}\|\theta\|_4^4 \|\theta\|^{2m-4}).$$

The last term is of a smaller order of the standard deviation of $\widehat{\chi}_{gc}^{(m)}$ and is thus negligible. Theorem 3.1 solidifies what we mentioned in Section 2.1 and is proved in Section 6.

**Theorem 3.2** *Consider the DCMM model* (1) *where* (7)-(9) *hold. As* $n \to \infty$, *for* $m = 3, 4$, *under either* $H_0^{(n)}$ *or* $H_1^{(n)}$,

$$\sqrt{\frac{B_{n,m}}{2m}} \cdot \widehat{C}_m^{-1/2}\left[\widehat{\chi}_{gc}^{(m)} - \chi_{gc}^{(m)}\right] \xrightarrow{d} N(0, 1).$$

The proof for other fixed $m$ is similar, but significantly more tedious so we leave it as future work. Theorem 3.2 is proved in the supplemental material. Compared to existing literature, our theorems are for a much broader setting where existing works have very little theory and understanding.

**Remark 6**. Conditions (8)-(9) are only for $H_1^{(n)}$ since they naturally hold under $H_0^{(n)}$; the conditions are only mild. Condition (7) is also only mild. Take the DCMM-RDP model for example (see Remark 1): $\|\theta\| \asymp \sqrt{n}\alpha_n$ and $\|\theta\|_3 \asymp n^{1/3}\alpha_n$, so Condition (7) requires $n^{-1/2} \ll \alpha_n \ll n^{-1/3}$. The case $\alpha_n \gg n^{-1/3}$ corresponds to the "strong signal" case, the analysis of which is different and we leave it to the forthcoming manuscript.

### 3.1. Testing Power

By Theorems 3.1-3.2, we expect to see that in distribution,

$$\sqrt{\frac{B_{n,m}}{2m}} \cdot \widehat{C}_m^{-1/2}\widehat{\chi}_{gc}^{(m)} \approx N\left(\sqrt{\frac{B_{n,m}}{2m}} C_m^{-1/2}\chi_{gc}^{(m)}, \ 1\right).$$

Motivated by Theorem 3.1 and equation (4), we introduce a proxy of $\sqrt{\frac{B_{n,m}}{2m}} \cdot C_m^{-1/2}\chi_{gc}^{(m)}$, defined as

$$\delta_{gc}^{(m)} = \frac{(2m)^{-1/2}}{\sqrt{\sum_{k=1}^K \lambda_k^m}}\left[\sum_{k=1}^K \lambda_k^m - \left(\frac{\sum_{k=1}^K (\eta, \xi_k)^2 \lambda_k^{m-1}}{\sum_{k=1}^K (\eta, \xi_k)^2 \lambda_k^{m-2}}\right)^m\right], \tag{11}$$

and we expect to see that in distribution,

$$\sqrt{\frac{B_{n,m}}{2m}} \cdot \widehat{C}_m^{-1/2}\widehat{\chi}_{gc}^{(m)} \approx N(\delta_{gc}^{(m)}, 1). \tag{12}$$

It is noteworthy that under $H_0^{(n)}$, $\delta_{gc}^{(m)} = 0$.

Fixing $0 < \alpha < 1$, let $z_\alpha$ be the $(1-\alpha)$-quantile of $N(0, 1)$. Consider the Graphlet Count (GC) test where we

$$\text{reject } H_0^{(n)} \iff \sqrt{\frac{B_{n,m}}{2m}} \cdot \widehat{C}_m^{-1/2}\widehat{\chi}_{gc}^{(m)} > z_\alpha. \tag{13}$$

The theorem below is proved in the supplemental material.

**Theorem 3.3** *Consider the DCMM model* (1) *where* (7)-(9) *hold. As* $n \to \infty$, *for* $m = 3, 4$, *the level and the power of the Graphlet Count test are respectively* $\alpha + o(1)$ *and* $\Phi(\delta_{gc}^{(m)} - z_\alpha) + o(1)$. *Moreover, if* $\delta_{gc}^{(m)} \to \infty$ *as* $n \to \infty$, *then the power* $\to 1$.

### 3.2. Comparison of $\widehat{\chi}_{gc}^{(3)}$ and $\widehat{\chi}_{gc}^{(4)}$

One of the key messages is that, $\widehat{\chi}_{gc}^{(3)}$ may be powerless for some non-null cases, even when the "signals" are strong and the test is relatively easy. In comparison, $\widehat{\chi}_{gc}^{(4)}$ has successfully avoided such a pitfall. Note that translated to our terms, the EZ test by Gao and Lafferty (2017) is $\widehat{\chi}_{gc}^{(3)}$.

In detail, by Remark 3, $\lambda_1$ is positive and has the largest absolute value among all $\lambda_k$'s, so

$$\left[\sum_{k=1}^K (\eta, \xi_k)^2 \lambda_k^{m-1}\right]/\left[\sum_{k=1}^K (\eta, \xi_k)^2 \lambda_k^{m-2}\right] \leq \lambda_1. \tag{14}$$

It follows that under $H_1^{(n)}$

$$\delta_{gc}^{(m)} \geq \left(\sum_{k=2}^K \lambda_k^m\right)/\left[2m \sum_{k=1}^K \lambda_k^m\right]^{1/2}, \tag{15}$$

where "=" is achievable; see Section 3.3 for examples.

It can be shown that $\sum_{k=1}^{K} \lambda_k^m > 0$, no matter $m$ is odd or even. The numerator $\sum_{k=2}^{K} \lambda_k^m$, however, is more tricky.

- When $m$ is even, the numerator of (15) is positive.
- When $m$ is odd, the numerator of (15) can be negative or 0. In fact, $\delta_{gc}^{(m)}$ may be 0 under some $H_1^{(n)}$, even when signals are strong; see Section 3.3 for an example. If additionally $P$ (and so $G^{1/2}PG^{1/2}$) is positive definite, then $\delta_{gc}^{(m)}$ is positive, either $m$ is odd or even.

**Corollary 3.1** *Consider the DCMM model (1) where (7)-(9) hold and the Graphlet Count test (13). As $n \to \infty$.*

- *When $m = 3$, for some configurations of the non-null case, $\delta_{gc}^{(3)} = 0$ and the power of the test is $\alpha + o(1)$. If additionally the $K \times K$ matrix $P$ is positive definite, then there is a constant $c_3 > 0$ such that $\delta_{gc}^{(3)} \geq c_3 \|\theta\|^3$ and so the power of the test $\gtrsim \Phi(c_3\|\theta\|^3 - z_\alpha)$, which tends to 1 since $\|\theta\| \to \infty$ in our setting.*

- *If $m = 4$, then there is a constant $c_4 > 0$ such that $\delta_{gc}^{(4)} \geq c_4 \|\theta\|^4$ and the power of the test $\gtrsim \Phi(c_4\|\theta\|^4 - z_\alpha)$, which tends to 1 as $\|\theta\| \to \infty$ in our setting.*

See the supplement for the proof. In comparison, $\widehat{\chi}_{gc}^{(3)}$ has a two-fold disadvantage: it may lose power in some non-null configurations even when $\|\theta\|$ is large, and as $\|\theta\| \to \infty$, its power grows to 1 in a speed slower than that of $\widehat{\chi}_{gc}^{(4)}$.

**Remark 7**. In the most subtle case where $\|\theta\| \asymp 1$, for any $m \geq 3$, $\widehat{\chi}_{gc}^{(m)}$ has a non-trivial power since the signal to noise ratio $\delta_{gc}^{(m)} \asymp \|\theta\|^m$. The form is reminiscent of the results on likelihood ratio (Mossel et al., 2015) which is unfortunately only for SBM (much narrower than DCMM).

**Remark 8**. The computational complexity of the GC test is $O(nd^2)$ for $m = 3$ and $O(nd^3)$ for $m = 4$ (Schank & Wagner, 2005), where $d$ is the maximum degree. Many large networks are reasonably sparse where $d \ll n$, and the complexity is reasonably modest in such cases.

**Remark 9**. Our work is connected to Maugis et al. (2017), which characterizes the expected number of closed walks. However, their work does not study the expected number non-closed walks, and the standard deviations of close and non-closed walks, so how to apply their results to our setting is unclear. Our work is also closed to the notion of clustering coefficient (Holland & Leinhardt, 1971; Watts & Strogatz, 1998), which in our notation equals to $3\widehat{C}_3/\widehat{L}_2$. To use this as a test, the challenge is how to normalize the statistic properly so the limiting distribution is more tractable. Also, the test may lose power in some settings. In Section 3.3, we provide an example where $C_3/L_2 = \lambda_1 + (\sum_{k=2}^{K} \lambda_k^3)/\lambda_1^2$, so when $\sum_{k=2}^{K} \lambda_k^3 = 0$, asymptotically, the test is powerless while the GC test may still have good power.

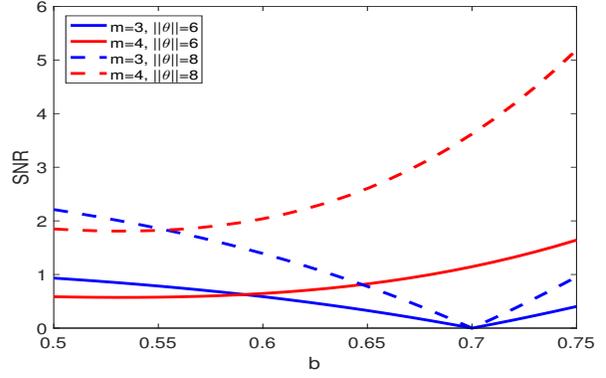

*Figure 1.* Plot of $\delta_{gc}^{(m)}$ in the setting of Section 3.3 $((K,a) = (10, .25))$. When $b = .7$, $\delta_{gc}^{(3)} = 0$ even when $\|\theta\|$ is very large, and the EZ test is powerless. In comparison, $\widehat{\chi}_{gc}^{(4)}$ has overcome such a pitfall. Also, $\widehat{\chi}_{gc}^{(4)}$ is more powerful when $\|\theta\|$ is large.

### 3.3. An Example

It is instructive to consider an example where $\delta_{gc}^{(m)}$ can be further simplified. Consider a setting where

- the DCMM-RDP model holds (see Remark 1), i.e., $(\theta_i/\alpha_n) \overset{i.i.d.}{\sim} f$, where the $2nd$ moment of $f$ is 1,
- all $\pi_i$'s are degenerate, dividing to $K$ equal-size communities,
- the rows of $P$ have an equal sum.

It follows that approximately: (a). $\|\theta\| = \sqrt{n}\alpha_n$, (b). $G \propto$ the $K \times K$ identity matrix, so $\xi_1$ (the first eigenvector of $G^{1/2}PG^{1/2}$) is approximately proportional to the vector of ones $1_K$. (c). $\eta = G^{-1/2}\Pi'\Theta 1_n \propto 1_K$, due to the random model for $\theta_i$. Therefore, $(\eta, \xi_k) = (\xi_1, \xi_k)$, which equals to 1 if $k = 1$ and 0 otherwise, and so by basic linear algebra,

$$\delta_{gc}^{(m)} = \frac{(2m)^{-1/2}}{\sqrt{\sum_{k=1}^{K}\lambda_k^m}}\left[\sum_{k=1}^{K}\lambda_k^m - \lambda_1^m\right] = \frac{\sum_{k=2}^{K}\lambda_k^m}{\sqrt{2m\sum_{k=1}^{K}\lambda_k^m}}.$$

We now consider a setting where we can spell out $\lambda_k$ more explicitly. Suppose $K$ is even and the $K \times K$ matrix $P$ calibrating the community structure is a $2 \times 2$ block-wise matrix having the form of $P = \begin{bmatrix} D & C \\ C & D \end{bmatrix}$, where $C, D \in \mathbb{R}^{K/2,K/2}$ and

$$D = \begin{bmatrix} 1 & a & \cdots & a \\ \vdots & & \ddots & \vdots \\ a & a & \cdots & 1 \end{bmatrix} \quad \text{and} \quad C = \begin{bmatrix} b & b & \cdots & b \\ \vdots & & \ddots & \vdots \\ b & b & \cdots & b \end{bmatrix}.$$

In this case, $\lambda_1 = \frac{n\alpha_n^2}{K}\{(1-a) + \frac{K}{2}(a+b)\}$, and the other $(K-1)$ eigenvalues are (which one is $\lambda_2$ depends on $(a,b)$)

$$\frac{n\alpha_n^2}{K}[(1-a) + \frac{K}{2}(a-b)], \frac{n\alpha_n^2}{K}(1-a), \ldots, \frac{n\alpha_n^2}{K}(1-a).$$

If we let $A_K(a,b) = (1-a) + \frac{K}{2}(a-b)$ and $B_K(a,b) = (1-a) + \frac{K}{2}(a+b)$, recalling $\|\theta\| = \sqrt{n}\alpha_n$, then

$$\delta_{gc}^{(3)} = \frac{(K^{-\frac{3}{2}}\|\theta\|^3) \cdot [A_K^3(a,b) + (K-2)(1-a)^3]}{\sqrt{[A_K^3(a,b) + B_K^3(a,b) + (K-2)(1-a)^3]}},$$

$$\delta_{gc}^{(4)} = \frac{(K^{-2}\|\theta\|^4) \cdot [A_K^4(a,b) + (K-2)(1-a)^4]}{\sqrt{A_K^4(a,b) + B_K^4(a,b) + (K-2)(1-a)^4}}.$$

Note that $\delta_{gc}^{(4)}$ is always positive, but $\delta_{gc}^{(3)}$ can be either positive or negative, and in particular,

$\delta_{gc}^{(3)} = 0$ when $b = w + (1-w)a$, where $w = \frac{1+(K-2)^{1/3}}{K/2}$.

Figure 1 compares $\delta_{gc}^{(3)}$ and $\delta_{gc}^{(4)}$ for $(K,a) = (10, .25)$ and various $(b, \|\theta\|)$. Regardless of $\|\theta\|$, $\delta_{gc}^{(3)}$ gets small when $b$ is close to 0.7. However, $\delta_{gc}^{(4)}$ has no such issue.

### 3.4. The Lower Bound

The lower bound is not discussed here but is studied in the extended version (Jin et al., 2018a), where we show that under DCMM, if $\|\theta\| \leq c$ for a sufficiently small constant, then the risk (sum of type-I and type-II errors) of any test converges to 1 as $n \to \infty$. See also Massoulié (2014), Mossel (2015), Abbe & Sandon (2016), Gao & Lafferty (2017). Note by Corollary 3.1, if the level $\alpha \to 0$, then the risk of the GC test $\to 0$ as $\|\theta\| \to \infty$. A closely related work is Jin and Ke (2017). Under the DCMM and the assumption of $\theta_{\max} \leq C\theta_{\min}$, they showed that when $\|\theta\| \not\to \infty$ it is impossible to successfully estimate the mixed memberships.

## 4. Simulations

We investigate $\widehat{\chi}_{gc}^{(m)}$ for $m = 3, 4$. Recall that when $m = 3$, it coincides with the EZ test (Gao & Lafferty, 2017). The methods (Bickel & Sarkar, 2016; Lei, 2016; Banerjee & Ma, 2017; Wang & Bickel, 2017) are for SBM, which do not apply to our settings and so we skip them for study.

**Experiment 1** (checking for asymptotic normality). Fixing $n = 200$, we consider a null setting where $\theta_i$'s are from $\sqrt{10}\theta_i \stackrel{iid}{\sim} Pareto(4, 0.375)$;[1] (note: severe degree heterogeneity!) We also consider an alternative case where $\theta_i$'s are from $\sqrt{2}\theta_i \stackrel{iid}{\sim} Pareto(4, 0.375)$, $K = 3$, and $P$ is the matrix with unit diagonals and all off-diagonals are $1/3$. Among the 200 nodes, 180 are pure with 60 in each community, and the remaining 20 nodes have a mixed membership $(1/3, 1/3, 1/3)$. The results of 500 repetitions are in Figure 2, which suggest that the claimed asymptotic normality is valid even for relatively small $n$.

**Experiment 2** (power comparison). Fix $(n, K) = (300, 10)$. All nodes are pure with 30 in each commu-nity. For $(a,b)$ and $h > 0$, let the matrix $P$ be the same as in Section 3.3. Set $\theta_i = (h/\|\tilde{\theta}\|)\tilde{\theta}_i$, where $\tilde{\theta}_i \stackrel{iid}{\sim} Pareto(4, 0.375)$; we note that $\|\theta\| = h$.

---
[1] In $Pareto(\alpha, x_m)$, $\alpha$ is for shape, $x_m$ is for scale.

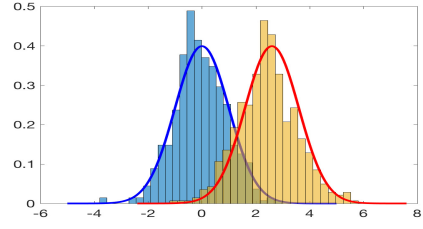

*Figure 2.* Histograms of $\sqrt{(B_{n,4}/8)} \cdot \widehat{C}_4^{-1/2}\widehat{\chi}_{gc}^{(4)}$ for the null hypothesis (light blue) and the alternative hypothesis (yellow). The blue and red curves are densities of $N(0,1)$ and $N(\delta_{gc}^{(4)}, 1)$, respectively, which support our results on asymptotical normality.

For $\|\theta\|$ ranging in $\{5, 6, \ldots, 10\}$, we consider three different settings for $(a,b)$, (i)-(iii). See Table 1 for the results. In (i), $(a,b) = (.15, .52)$ and the second eigenvalue $\lambda_2$ of $G^{1/2}PG^{1/2}$ is moderately large, and $\widehat{\chi}_{gc}^{(4)}$ uniformly outperforms $\widehat{\chi}_{gc}^{(3)}$ (the EZ test). In (ii), $(a,b) = (.30, .54)$, $\lambda_2$ is relatively small and the testing problem is more challenging than (i). In this case, $\widehat{\chi}_{gc}^{(4)}$ performs better again. In (iii), we investigate a case where $\delta_{gc}^{(3)} = 0$ and see how $\widehat{\chi}_{gc}^{(3)}$ deals with this most challenging case (the case poses no challenge to $\widehat{\chi}_{gc}^{(4)}$ as $\delta_{gc}^{(4)}$ is always $> 0$; see Section 3.3). In this case, $\widehat{\chi}_{gc}^{(3)}$ loses power, and significantly underperforms $\widehat{\chi}_{gc}^{(4)}$.

These results are consistent with our theoretical results, especially those of Section 3.3.

*Table 1.* Powers of $\widehat{\chi}_{gc}^{(4)}$ (GC) and $\widehat{\chi}_{gc}^{(3)}$ (EZ).

| $(a,b)$ | $\|\theta\|$ | 5 | 6 | 7 | 8 | 9 | 10 |
|---|---|---|---|---|---|---|---|
| (.15, .52) | GC | .13 | .46 | .77 | .99 | 1.0 | 1.0 |
|  | EZ | .07 | .17 | .28 | .64 | .93 | .99 |
| (.30, .54) | GC | .11 | .12 | .27 | .55 | .85 | .99 |
|  | EZ | .07 | .10 | .22 | .25 | .51 | .78 |
| (.15, .66) | GC | .09 | .27 | .72 | .94 | 1.0 | 1.0 |
|  | EZ | .08 | .05 | .07 | .06 | .16 | .43 |

## 5. Application to a Football Network

In the college football network (Girvan & Newman, 2002), each node is a Division I-A college team and two nodes have an edge if and only if they played $\geq 1$ games during the Fall 2000 season. There are a total of 115 teams. Except for 5 "independent" teams, all teams are manually divided into 11 conferences for administration purposes; we treat these "manually labeled communities" as the ground truth.

First, we consider a relatively easy setting where we test whether the whole network has only 1 community or has multiple communities. As expected, for both $m = 3$ and $4$, our test $\widehat{\chi}_{gc}^{(m)}$ rejects the null with extremely small $p$-values.

Next, we consider a more subtle problem, where for each of the 11 *manually labeled communities* aforementioned, we test whether it can't be further divided into multiple communities (null) or it can (alternative). The results of all 11 testing settings are in Table 2.

For the first 10 test settings, despite some differences in the $p$-values, two tests, $\widehat{\chi}_{gc}^{(3)}$ and $\widehat{\chi}_{gc}^{(4)}$, agree with each other and both accept the null. For the last setting (corresponding to the Western Athletic Conference (WAC)), however, $\widehat{\chi}_{gc}^{(4)}$ votes for rejection and $\widehat{\chi}_{gc}^{(3)}$ votes for acceptance. It turns out that one team ("BioseState") in the WAC is an outlier, which did not play any game in the data range. After removing the outlier, both tests vote for acceptance. These results are consistent with the ground truth, suggesting (a) both tests yield reasonable testing results even for small-size networks, and (b) $\widehat{\chi}_{gc}^{(4)}$ is more effective in detecting outliers.

*Table 2.* The 11 testing results (each corresponds to a conference). Column 3-4: test scores $\sqrt{B_{n,m}/(2m)}\widehat{C}_m^{-1/2}\widehat{\chi}_{gc}^{(m)}$ and corresponding $p$-values with $m = 3$. Column 5-6: same but for $m = 4$.

| Conference | size | score | p-value | score | p-value |
|---|---|---|---|---|---|
| Atlantic Coast | 9 | 0.00 | 1.00 | 0.00 | 0.50 |
| Big East | 8 | 0.00 | 1.00 | 0.00 | 0.50 |
| Big Ten | 11 | -0.07 | 0.94 | -0.31 | 0.62 |
| Big Twelve | 12 | -0.02 | 0.98 | -0.48 | 0.68 |
| Conference USA | 10 | 0.26 | 0.80 | 1.23 | 0.11 |
| Mid-American | 13 | 0.65 | 0.51 | 0.24 | 0.41 |
| Mountain West | 8 | 0.00 | 1.00 | 0.00 | 0.50 |
| Pacific Ten | 10 | -0.04 | 0.97 | -0.19 | 0.58 |
| Southeastern | 12 | -0.06 | 0.95 | -0.40 | 0.65 |
| Sun Belt | 7 | 1.48 | 0.14 | 1.06 | 0.15 |
| Western Athletic | 10 | 0.51 | 0.61 | **2.48** | **0.01** |

## 6. Proof of Theorem 3.1

First, we show that for any $m \geq 3$,

$$B_{n,m}C_m = \sum_{k=1}^K \lambda_k^m + O\big(\|\theta\|_4^4 \|\theta\|^{2m-4}\big), \quad (16)$$

and for any $m \geq 1$, $B_{n,m+1}L_m$ equals to

$$\sum_{k=1}^K \lambda_k^m (\eta'\xi_k)^2 + O\big(\|\theta\|_1^2 \|\theta\|_4^4 \|\theta\|^{2m-6}\big). \quad (17)$$

Consider (16). By definition, $B_{n,m}C_m = B_{n,m}\mathbb{E}[\widehat{C}_m] = \sum_{i_1,\ldots,i_m} \mathbb{E}[A_{i_1i_2}\cdots A_{i_mi_1}] = \sum_{i_1,\ldots,i_m} \Omega_{i_1i_2}\cdots \Omega_{i_mi_1}$, where the sum is over distinct indices $i_1, \ldots, i_m$. As a result,

$$B_{n,m}C_m = \operatorname{tr}(\Omega^m) - \sum_{\substack{\text{non-distinct} \\ i_1,\ldots,i_m}} \Omega_{i_1i_2}\Omega_{i_2i_3}\cdots \Omega_{i_mi_1}.$$

We calculate the term $\operatorname{tr}(\Omega^m)$. From the DCMM model, $\Omega = \Theta\Pi P \Pi'\Theta$. It follows that

$$\Omega^m = \Theta\Pi P(\Pi'\Theta^2\Pi)P(\Pi'\Theta^2\Pi)\cdots P(\Pi'\Theta^2\Pi)P\Pi'\Theta$$

$$= \Theta\Pi(PGPG\cdots PGP)\Pi'\Theta$$

$$= (\Theta\Pi G^{-1/2})(G^{1/2}PG^{1/2})^m (G^{-1/2}\Pi'\Theta).$$

For any matrices $A$ and $B$, $\operatorname{tr}(AB) = \operatorname{tr}(BA)$. As a result,

$$\operatorname{tr}(\Omega^m) = \operatorname{tr}\big[(P^{1/2}GP^{1/2})^m (G^{-1/2}\Pi'\Theta)(\Theta\Pi G^{-1/2})\big]$$
$$= \operatorname{tr}\big[(P^{1/2}GP^{1/2})^m\big] = \sum_k \lambda_k^m. \quad (18)$$

We then bound the remainder term. Note that $\Omega_{ij} = \theta_i\theta_j(\pi_i'P\pi_j) \leq C\theta_i\theta_j$, where the last inequality is from Condition (9). Hence,

$$\sum_{\substack{\text{non-distinct} \\ i_1,\ldots,i_m}} \Omega_{i_1i_2}\Omega_{i_2i_3}\cdots \Omega_{i_mi_1} \leq \sum_{\substack{\text{non-distinct} \\ i_1,\ldots,i_m}} C\theta_{i_1}^2\theta_{i_2}^2\cdots\theta_{i_m}^2$$

$$\leq \sum_{i_1,\ldots,i_{m-1}} C\theta_{i_1}^4\theta_{i_2}^2\cdots\theta_{i_{m-1}}^2 \leq C\|\theta\|_4^4\|\theta\|^{2(m-2)}.$$

Combining the above gives (16).

Consider (17). Similarly, we have

$$B_{n,m+1}L_m = 1_n'\Omega^m 1_n - \sum_{\substack{\text{non-distinct} \\ i_1,\ldots,i_{m+1}}} \Omega_{i_1i_2}\Omega_{i_2i_3}\cdots\Omega_{i_m i_{m+1}}.$$

Since $\Omega = \Theta\Pi P\Pi'\Theta$, it follows that

$$1_n'\Omega^m 1_n = 1_n'\Theta\Pi P(\Pi'\Theta^2\Pi)P(\Pi'\Theta^2\Pi)\cdots P\Pi'\Theta 1_n$$
$$= 1_n'\Theta\Pi(PGPG\cdots P)\Pi'\Theta 1_n$$
$$= 1_n'\Theta\Pi G^{-1/2}(G^{1/2}PG^{1/2})^m G^{-1/2}\Pi'\Theta 1_n$$
$$= \eta'(G^{1/2}PG^{1/2})^m \eta.$$

The eigen-decomposition $G^{1/2}PG^{1/2} = \sum_{k=1}^K \lambda_k\xi_k\xi_k'$ implies that $(G^{1/2}PG^{1/2})^m = \sum_{k=1}^K \lambda_k^m \xi_k\xi_k'$. As a result,

$$1_n'\Omega^m 1_n = \eta'\big[\sum_{k=1}^K \lambda_k^m \xi_k\xi_k'\big]\eta = \sum_{k=1}^K \lambda_k^m (\eta'\xi_k)^2. \quad (19)$$

We then bound the remainder term. Since $\Omega_{ij} \leq C\theta_i\theta_j$, $\Omega_{i_1i_2}\cdots\Omega_{i_m i_{m+1}} \leq C\theta_{i_1}\theta_{i_{m+1}}\theta_{i_2}^2\cdots\theta_{i_m}^2$. It follows that

$$\sum_{\substack{\text{non-distinct} \\ i_1,\ldots,i_{m+1}}} \Omega_{i_1i_2}\Omega_{i_2i_3}\cdots\Omega_{i_m i_{m+1}}$$

$$\leq \Big(\sum_{i_1 = i_{m+1}} + \sum_{i_2 = i_{m+1}} + \sum_{\substack{\text{non-distinct} \\ i_2,\ldots,i_m}}\Big) C\theta_{i_1}\theta_{i_{m+1}}\theta_{i_2}^2\cdots\theta_{i_m}^2$$

$$\leq \sum_{i_1,\ldots,i_m} C\theta_{i_1}^2\cdots\theta_{i_m}^2 + \sum_{i_1,\ldots,i_m} C\theta_{i_1}\theta_{i_2}^3\theta_{i_3}^2\cdots\theta_{i_m}^2$$

$$+ \sum_{i_1,\ldots,i_{m-1},i_{m+1}} \theta_{i_1}\theta_{i_{m+1}}\theta_{i_2}^4\theta_{i_3}^2\cdots\theta_{i_{m-1}}^2$$

$$\leq C\Big[\|\theta\|^{2m} + \|\theta\|_1\|\theta\|_3^3\|\theta\|^{2(m-2)} + \|\theta\|_1^2\|\theta\|_4^4\|\theta\|^{2(m-3)}\Big]$$

$$\leq C\|\theta\|^{2(m-3)}\big(\|\theta\|^6 + \|\theta\|_1\|\theta\|_3^3\|\theta\|^2 + \|\theta\|_1^2\|\theta\|_4^4\big).$$

We need to compare the three terms in the brackets. First, applying Holder's inequality with $p = 3$ and $q = 3/2$, we have $\sum_i \theta_i^2 = \sum_i \theta_i^{\frac{4}{3}}\theta_i^{\frac{2}{3}} \leq (\sum_i \theta_i^{\frac{4p}{3}})^{\frac{1}{p}}(\sum_i \theta_i^{\frac{2q}{3}})^{\frac{1}{q}}$. It implies $\|\theta\|^2 \leq \|\theta\|_4^{4/3}\|\theta\|_1^{2/3}$. As a result,
$$\|\theta\|^6 \leq \|\theta\|_4^4\|\theta\|_1^2.$$
This means the first term above is dominated by the last term. Second, by Cauchy-Schwartz inequality, $\sum_i \theta_i^3 \leq (\sum_i \theta_i^2)^{1/2}(\sum_i \theta_i^4)^{1/2}$, which means $\|\theta\|_3^3 \leq \|\theta\|\|\theta\|_4^2$. As a result, $\|\theta\|_1\|\theta\|_3^3\|\theta\|^2 \leq \|\theta\|_1\|\theta\|_4^2\|\theta\|^3$. Furthermore, we have proved $\|\theta\|^3 \leq \|\theta\|_4^2\|\theta\|_1$. Combining the above gives $\|\theta\|_1\|\theta\|_3^3\|\theta\|^2 \leq \|\theta\|_1^2\|\theta\|_4^4$. Hence, the second term is dominated by the last term. In summary, the remainder term is $O(\|\theta\|_1^2\|\theta\|_4^4\|\theta\|^{2(m-3)})$. This proves (17).

Next, we use (16)-(17) to show the claim. Write for short
$$\chi_{gc,0}^{(m)} = \frac{1}{B_{n,m}}\left\{\sum_k \lambda_k^m - \left[\frac{\sum_k(\eta, \xi_k)^2\lambda_k^{m-1}}{\sum_k(\eta, \xi_k)^2\lambda_k^{m-2}}\right]^m\right\}.$$

Write $\widetilde{C}_m = B_{n,m}C_m$, $\widetilde{C}_m^0 = \mathrm{tr}(\Omega^m)$, $\widetilde{L}_m = B_{n,m+1}L_m$, and $\widetilde{L}_m^0 = 1_n'\Omega^m 1_n$, for all $m$. By definition and (18)-(19),
$$B_{n,m}\chi_{gc,0}^{(m)} = \widetilde{C}_m^0 - (\widetilde{L}_{m-1}^0/\widetilde{L}_{m-2}^0)^m,$$
$$B_{n,m}\chi_{gc}^{(m)} = B_{n,m}[C_m - (L_{m-1}/L_{m-2})^m]$$
$$= \widetilde{C}_m - \frac{(B_{n,m-1})^m}{(B_{n,m})^{m-1}}\cdot(\widetilde{L}_{m-1}/\widetilde{L}_{m-2})^m.$$

As a result,
$$B_{n,m}|\chi_{gc}^{(m)} - \chi_{gc,0}^{(m)}|$$
$$\leq |\widetilde{C}_m - \widetilde{C}_m^0| + \left|\frac{(\widetilde{L}_{m-1})^m}{(\widetilde{L}_{m-2})^m} - \frac{(\widetilde{L}_{m-1}^0)^m}{(\widetilde{L}_{m-2}^0)^m}\right|$$
$$+ \frac{(\widetilde{L}_{m-1})^m}{(\widetilde{L}_{m-2})^m}\left|\frac{(B_{n,m-1})^m}{(B_{n,m})^{m-1}} - 1\right|$$
$$\equiv I_1 + I_2 + I_3. \qquad (20)$$

We now bound these three terms. By (16)-(17),
$$|\widetilde{C}_m - \widetilde{C}_m^0| = O(\|\theta\|_4^4\|\theta\|^{2m-4}),$$
$$|\widetilde{L}_m - \widetilde{L}_m^0| = O(\|\theta\|_1^2\|\theta\|_4^4\|\theta\|^{2m-6}). \qquad (21)$$

Hence,
$$I_1 = O(\|\theta\|_4^4\|\theta\|^{2m-4}).$$

To bound $I_2$, we need the following lemma, which is proved in the supplemental material.

**Lemma 6.1** *Under conditions of Theorem 3.3, $|\lambda_k| \asymp \|\theta\|^2$ for $1 \leq k \leq K$, and $\max_{1\leq k\leq K} |\eta'\xi_k| \asymp \|\theta\|^{-1}\|\theta\|_1$.*

By (19), $\widetilde{L}_m^0 = \sum_k(\eta'\xi_k)^2\lambda_k^m$. For $m$ even, it then follows from Lemma 6.1 that
$$\widetilde{L}_m^0 \geq c\|\theta\|^{2m-2}\|\theta\|_1^2 \qquad (22)$$

for a constant $c > 0$. For $m$ odd, this is still true; see the proof of Lemma B.1 in the supplemental material. Additionally, by Cauchy-Schwarz inequality, $\|\theta\|^4 \leq n\|\theta\|_4^4$; hence, for $m \leq 3$, $|\widetilde{L}_m - \widetilde{L}_m^0|$ is negligible compared to the order of $\widetilde{L}_m^0$, so we also have $\widetilde{L}_m \geq c\|\theta\|^{2m-2}\|\theta\|_1^2$. Now,

$$\left|\frac{\widetilde{L}_{m-1}^0}{\widetilde{L}_{m-2}^0} - \frac{\widetilde{L}_{m-1}}{\widetilde{L}_{m-2}}\right| \leq \frac{|\widetilde{L}_{m-1} - \widetilde{L}_{m-1}^0|}{\widetilde{L}_{m-2}^0} + \frac{\widetilde{L}_{m-1}}{\widetilde{L}_{m-2}}\frac{|\widetilde{L}_{m-2} - \widetilde{L}_{m-2}^0|}{\widetilde{L}_{m-2}^0}$$
$$= O\Big(\frac{\|\theta\|_1^2\|\theta\|_4^4\|\theta\|^{2m-8}}{\|\theta\|^{2m-6}\|\theta\|_1^2}\Big) + O\Big(\|\theta\|^2\frac{\|\theta\|_1^2\|\theta\|_4^4\|\theta\|^{2m-10}}{\|\theta\|^{2m-6}\|\theta\|_1^2}\Big)$$
$$= O(\|\theta\|_4^4\|\theta\|^{-2}).$$

Since $|x^m - y^m| \leq C|x - y|(|x| + |y|)^{m-1}$,
$$I_2 \leq C\left|\frac{\widetilde{L}_{m-1}^0}{\widetilde{L}_{m-2}^0} - \frac{\widetilde{L}_{m-1}}{\widetilde{L}_{m-2}}\right|\cdot\left|\frac{\widetilde{L}_{m-1}^0}{\widetilde{L}_{m-2}^0}\right|^{m-1}$$
$$= O(\|\theta\|_4^4\|\theta\|^{-2}\cdot\|\theta\|^{2m-2}) = O(\|\theta\|_4^4\|\theta\|^{2m-4}).$$

Consider $I_3$. Note that $\frac{(B_{n,m-1})^m}{(B_{n,m})^{m-1}} = \frac{n(n-1)\cdots(n-m+2)}{(n-m+1)^{m-1}} = \prod_{j=1}^{m-1}(1 + \frac{j}{n-m+1}) = 1 + O(\frac{1}{n})$. So,
$$I_3 = O(\|\theta\|^{2m}\cdot n^{-1}) = O(\|\theta\|_4^4\|\theta\|^{2m-4}),$$

where we have used the universal inequality $\|\theta\|^4 \leq n\|\theta\|_4^4$.

We plug the above results into (20) and note that $B_{n,m} \sim n^m$. The claim follows immediately. $\square$

## 7. Conclusion

We consider a hard testing problem in the rather general DCMM model where the challenge is severe degree heterogeneity. We discover a systematic way to cancel the effects of degree heterogeneity, and propose a family of tests, with careful analysis and numerical support. Compared to literature, our tests have competitive powers and are applicable in much broader settings. Our theory is also for very broad settings where existing works have very limited understanding. We point out an unappealing feature of the EZ test (Gao & Lafferty, 2017), and shows a new test in our family has successfully overcome the problem that the EZ test faces.

In our theorems, we require $K$ to be fixed, but the results continue to hold if $K \to \infty$ reasonably slowly. We also require the singular values of $P$ are in the same order, but this is mostly for simplicity in presentation and can be replaced by weaker conditions. We also assume $\|\theta\|_3 \to 0$. The case $\|\theta\|_3 \to \infty$ is related to the "dense network" case, the analysis of which can be done but is different and we leave it as future work.

## A. An Alternative Expression of the GC Test Statistics

We rewrite the test statistic $\widehat{\chi}_{gc}$ (as well as $\widehat{L}_2$, $\widehat{L}_3$ and $\widehat{C}_4$) explicitly as a function of the adjacency matrix $A$. The following proposition is proved in Section D.4.

**Proposition A.1** *The following are true:*

$$\widehat{L}_2 = \frac{1}{6\binom{n}{3}}\big[1'A^2 1 - \mathrm{tr}(A^2)\big],$$

$$\widehat{L}_3 = \frac{1}{24\binom{n}{4}}\big[1'A^3 1 - 2(1'A^2 1) + 1'A1 - \mathrm{tr}(A^3)\big],$$

*and*

$$\widehat{C}_4 = \frac{1}{24\binom{n}{4}}\big[\mathrm{tr}(A^4) - 2(1'A^2 1) + 1'A1\big].$$

*Furthermore,*

$$\widehat{\chi}_{gc} = \frac{\big[\mathrm{tr}(A^4) - 2(1'A^2 1) + 1'A1\big]}{n(n-1)(n-2)(n-3)}$$
$$- \frac{1}{(n-3)^4}\left[\frac{1'A^3 1 - 2(1'A^2 1) + 1'A1 - \mathrm{tr}(A^3)}{1'A^2 1 - \mathrm{tr}(A^2)}\right]^4.$$

## B. Proof of Theorem 3.2

We prove the case $m=4$. The case of $m=3$ is similar and thus omitted. From now on, we omit the superscripts "(4)" in all related quantities (e.g., we write $\delta_{gc}^{(4)}$ as $\delta_{gc}$). Write

$$\frac{\sqrt{3\binom{n}{4}}}{\sqrt{\widehat{C}_4}}(\widehat{\chi}_{gc} - \chi_{gc}) = \sqrt{\frac{C_4}{\widehat{C}_4}} \cdot (I + II) \quad (23)$$

where

$$I = \frac{\sqrt{3\binom{n}{4}}}{\sqrt{C_4}}(\widehat{C}_4 - C_4), \quad II = -\frac{\sqrt{3\binom{n}{4}}}{\sqrt{C_4}}\left[\Big(\frac{\widehat{L}_3}{\widehat{L}_2}\Big)^4 - \Big(\frac{L_3}{L_2}\Big)^4\right].$$

Using the Slutsky's theorem, it suffices to show that

$$\widehat{C}_4/C_4 \xrightarrow{p} 1, \quad (24)$$

$$I \xrightarrow{d} N(0,1), \quad (25)$$

and

$$II \xrightarrow{p} 0, \quad (26)$$

The following lemma is useful, and its proof can be found in Section D.

**Lemma B.1** *Under the assumptions of Theorem 3.2,*

$$C_4 \asymp n^{-4}\|\theta\|^8, \quad L_2 \asymp n^{-3}\|\theta\|_1^2\|\theta\|^2,$$

$$L_3 \asymp n^{-4}\|\theta\|_1^2\|\theta\|^4.$$

*Moreover,*

$$\mathrm{Var}(\widehat{C}_4) \leq Cn^{-8}\|\theta\|^8, \quad \mathrm{Var}(\widehat{L}_2) \leq Cn^{-6}\|\theta\|_1^3\|\theta\|_3^3,$$

$$\mathrm{Var}(\widehat{L}_3) \leq Cn^{-8}\|\theta\|_1^4\|\theta\|_3^6.$$

We now show (24)-(26). The proof of (25) is relatively long, so we prove it in the end.

First, we prove (24). Recall that $C_4 = E[\widehat{C}_4]$. By Lemma B.1,

$$\mathbb{E}[(\widehat{C}_4/C_4 - 1)^2] = C_4^{-2}\mathrm{Var}(\widehat{C}_4) = O(\|\theta\|^{-8}),$$

where the right hand side $\to 0$ as $\|\theta\| \to \infty$. The claim follows by elementary probability theory.

Second, we prove (26). Define $\widehat{L}_2^* = (\|\theta\|^2/n)\widehat{L}_2$ and $L_2^* = (\|\theta\|^2/n)L_2$. Using Lemma B.1, it follows from direct calculations that

$$L_3/L_2^* = O(1). \quad (27)$$

With these notations, we have

$$|II| = \frac{\sqrt{3\binom{n}{4}}}{\sqrt{C_4}} \cdot \frac{\|\theta\|^8}{n^4}\left|\Big(\frac{\widehat{L}_3}{\widehat{L}_2^*}\Big)^4 - \Big(\frac{L_3}{L_2^*}\Big)^4\right|$$

$$\leq C\|\theta\|^4\left|\Big(\frac{\widehat{L}_3}{\widehat{L}_2^*}\Big)^4 - \Big(\frac{L_3}{L_2^*}\Big)^4\right|,$$

where we have used $C_4 \asymp n^{-4}\|\theta\|^8$ in the second equality; see Lemma B.1. Note that for any $(x,y)$, $|x^4 - y^4| = |(x-y)(x^3 + x^2 y + xy^3 + y^3)| \leq 3|x-y| \cdot (|x| + |y|)^3$. It follows that

$$|II| \leq C \cdot \|\theta\|^4 |Z| \cdot \Big(\frac{L_3}{L_2^*} + |Z|\Big)^3,$$

where for short we write

$$Z = \frac{\widehat{L}_3}{\widehat{L}_2^*} - \frac{L_3}{L_2^*}.$$

Recall that $L_3/L_2^*$ is bounded. Therefore, to show (26), it suffices to show

$$\|\theta\|^4\Big(\frac{\widehat{L}_3}{\widehat{L}_2^*} - \frac{L_3}{L_2^*}\Big) \xrightarrow{p} 0. \quad (28)$$

Below, we show (28). Write the term on the left by

$$\frac{\|\theta\|^4}{L_2^*}(\widehat{L}_3 - L_3) + \|\theta\|^4 \frac{\widehat{L}_3}{\widehat{L}_2^* L_2^*}(L_2^* - \widehat{L}_2^*) \equiv II_a + II_b.$$

To show (28), it suffices to show

$$II_a \xrightarrow{p} 0. \quad (29)$$

and
$$II_b \xrightarrow{p} 0. \qquad (30)$$

Consider (29). Note that $L_3 = E[\widehat{L}_3]$. It follows from Lemma B.1 that

$$\mathrm{Var}(II_a) = \frac{\|\theta\|^8 \mathrm{Var}(\widehat{L}_3)}{(L_2^*)^2} \leq C \frac{\|\theta\|^8 \cdot n^{-8} \|\theta\|_1^4 \|\theta\|_3^6}{(n^{-4} \|\theta\|_1^2 \|\theta\|^4)^2}$$
$$\leq C \|\theta\|_3^6,$$

where the last term $\to 0$ for $\|\theta\|_3 \to 0$ as $n \to \infty$; this is due to equation (7) of (Jin et al., 2018b). By elementary probability, (29) follows.

Consider (30). To show the claim, we first show
$$\widehat{L}_2/L_2 \xrightarrow{p} 1, \qquad \widehat{L}_3/L_3 \xrightarrow{p} 1; \qquad (31)$$

as the proofs are similar, we only show the first one. By Lemma B.1, $\mathrm{Var}(\widehat{L}_2) = O(n^{-6} \|\theta\|_1^3 \|\theta\|_3^3)$ and $L_2 \asymp n^{-3} \|\theta\|_1^2 \|\theta\|^2$. Using $E[\widehat{L}_2] = L_2$, $\mathbb{E}[(\widehat{L}_2/L_2 - 1)^2] = L_2^{-2} \mathrm{Var}(\widehat{L}_2) \leq C(\|\theta\|_3^3)/(\|\theta\|_1 \|\theta\|^4))$, which $\leq C/\|\theta\|^2$ since $\|\theta\|_3^3 \leq \|\theta\|_1 \|\theta\|^2$. This shows (31).

Using (31) and recalling $L_3/L_2^* \leq C$ (see (27)), to show (30), it is sufficient to show
$$\|\theta\|^4 \frac{1}{(L_2^*)^2}(\widehat{L}_2^* - L_2^*) \xrightarrow{p} 0,$$

and since $\widehat{L}_2^*/L_2^* = \widehat{L}_2/L_2$, it is equivalent to show
$$\|\theta\|^4 \left(\frac{\widehat{L}_2}{L_2} - 1\right) \xrightarrow{p} 0. \qquad (32)$$

Last, we prove (25). We need some notations. Given 4 distinct nodes, there are 3 different possible cycles, denoted as $CC(i_1, i_2, i_3, i_4) = \{(i_1, i_2, i_3, i_4), (i_1, i_2, i_4, i_3), (i_1, i_3, i_2, i_4)\}$; moreover, for $B \subset \{1, 2, ..., n\}^4$, let $CC(B) = \cup_{(i_1,i_2,i_3,i_4) \in B} CC(i_1, i_2, i_3, i_4)$. For $1 \leq m \leq n$, let $I_m$ be the collection of $(i_1, i_2, i_3, i_4)$ such that $1 \leq i_1 < i_2 < i_3 < i_4 = m$. Write $\Omega_{ij}^* = \Omega_{ij}(1 - \Omega_{ij})$. Let $W = A - \Omega$. Define
$$S_{n,n} \equiv \frac{\sum_{CC(I_n)} W_{i_1 i_2} W_{i_2 i_3} W_{i_3 i_4} W_{i_4 i_1}}{\sqrt{\sum_{CC(I_n)} \Omega_{i_1 i_2}^* \Omega_{i_2 i_3}^* \Omega_{i_3 i_4}^* \Omega_{i_4 i_1}^*}}.$$

The following lemma is proved in Section D.

**Lemma B.2** *Under the conditions of Theorem 3.2,*
$$\frac{\sqrt{3\binom{n}{4}}}{\sqrt{C_4}}(\widehat{C}_4 - C_4) - S_{n,n} \xrightarrow{p} 0.$$

By Lemma B.2, to show (25), it suffices to show that
$$S_{n,n} \xrightarrow{d} N(0, 1). \qquad (33)$$

Below, we prove (33). For $1 \leq m \leq n$, define the $\sigma$-algebra $\mathcal{F}_{n,m} = \sigma(\{A_{ij}\}_{1 \leq i < j \leq m})$ and
$$X_{n,m} = S_{n,m} - S_{n,m-1},$$
where $S_{n,0} = 0$ and
$$S_{n,m} = \frac{\sum_{CC(I_m)} W_{i_1 i_2} W_{i_2 i_3} W_{i_3 i_4} W_{i_4 i_1}}{\sqrt{\sum_{CC(I_n)} \Omega_{i_1 i_2}^* \Omega_{i_2 i_3}^* \Omega_{i_3 i_4}^* \Omega_{i_4 i_1}^*}}, \quad 1 \leq m \leq n.$$

It is easy to see that $\mathbb{E}[S_{n,m}|\mathcal{F}_{n,m-1}] = S_{n,m-1}$. Hence, $\{X_{n,m}\}_{m=1}^n$ is a martingale difference sequence relative to the filtration $\{\mathcal{F}_{n,m}\}_{m=1}^n$, and $S_{n,n} = \sum_{m=1}^n X_{n,m}$. To show (33), we apply the martingale central limit theorem in (Hall & Heyde, 2014) and check:

(a) $\sum_{m=1}^n \mathbb{E}(X_{n,m}^2 | \mathcal{F}_{n,m-1}) \xrightarrow{p} 1$.

(b) $\sum_{m=1}^n \mathbb{E}(X_{n,m}^2 \mathbf{1}_{\{|X_{n,m}| > \epsilon\}} | \mathcal{F}_{n,m-1}) \xrightarrow{p} 0$, for any $\epsilon > 0$.

Note that once we have checked that both conditions (a) and (b) are satisfied, then by the martingale central limit theorem, $S_{n,n} \xrightarrow{d} N(0, 1)$. Combining it with Lemma B.2, we have proved (25).

It remains to check (a)-(b). For preparation, we first derive an alternative expression of $\mathbb{E}(X_{n,m}^2 | \mathcal{F}_{n,m-1})$ as (36) below. By definition,
$$X_{n,m} = \frac{1}{\sqrt{M_n}} \sum_{\sum_{CC(I_m) \setminus CC(I_{m-1})}} W_{i_1 i_2} W_{i_2 i_3} W_{i_3 i_4} W_{i_4 i_1},$$

where $M_n \equiv \sum_{CC(I_n)} \Omega_{i_1 i_2}^* \Omega_{i_2 i_3}^* \Omega_{i_3 i_4}^* \Omega_{i_4 i_1}^*$ and the summation is over all 4-cycles in $CC(I_m) \setminus CC(I_{m-1})$. Note that a cycle in $CC(I_m) \setminus CC(I_{m-1})$ has to include the node $m$. Hence, we can use the following way to get all such cycles: First, select 2 indices $(i, j)$ from $\{1, 2, ..., m-1\}$ and use them as the two neighboring nodes of $m$; second, select an index $k \in \{1, 2, ..., m-1\} \setminus \{i, j\}$ as the last node in the cycle. This allows us to write
$$X_{n,m} = \frac{1}{\sqrt{M_n}} \sum_{1 \leq i < j \leq m-1} W_{mi} W_{mj} \cdot Y_{(m-1)ij}, \quad (34)$$
where
$$Y_{(m-1)ij} = \sum_{1 \leq k \leq m-1, k \notin \{i,j\}} W_{ki} W_{kj}. \quad (35)$$

Conditioning on $\mathcal{F}_{n,m-1}$, $\{W_{mi} W_{mj}\}_{1 \leq i < j \leq m-1}$ are mutually uncorrelated and $Y_{(m-1)ij}$ is a constant. Hence, it follows from (34)-(35) that
$$\mathbb{E}(X_{n,m}^2 | \mathcal{F}_{n,m-1}) = \frac{1}{M_n} \sum_{1 \leq i < j \leq m-1} Y_{(m-1)ij}^2 \Omega_{mi}^* \Omega_{mj}^*. \qquad (36)$$

We now check (a). It suffices to show that

$$\mathbb{E}\Big[\sum_{m=1}^n \mathbb{E}(X_{n,m}^2|\mathcal{F}_{n,m-1})\Big] = 1, \tag{37}$$

and

$$\mathrm{Var}\Big(\sum_{m=1}^n \mathbb{E}(X_{n,m}^2|\mathcal{F}_{n,m-1})\Big) \to 0. \tag{38}$$

Consider (37). In the definition (35), the terms in the sum are (unconditionally) mutually uncorrelated. As a result,

$$\mathbb{E}[Y_{(m-1)ij}^2] = \sum_{k<m, k\notin\{i,j\}} \Omega_{ki}^* \Omega_{kj}^*.$$

It follows that

$$\mathbb{E}\Big[\sum_{m=1}^n \mathbb{E}(X_{n,m}^2|\mathcal{F}_{n,m-1})\Big]$$
$$= \frac{1}{M_n}\sum_{m=1}^n \sum_{1\le i<j\le m-1} \sum_{1\le k\le m-1, k\notin\{i,j\}} \Omega_{ki}^*\Omega_{kj}^*\Omega_{mi}^*\Omega_{mj}^*$$
$$= \frac{1}{M_n} \sum_{(m,i,j,k)\in CC(I_n)} \Omega_{mi}^*\Omega_{ik}^*\Omega_{kj}^*\Omega_{jm}^* = 1. \tag{39}$$

This proves (39).

Consider (38). We first decompose the random variable $\sum_{m=1}^n \mathbb{E}(X_{n,m}^2|\mathcal{F}_{n,m-1})$ into the sum of two parts, and then calculate its variance. By (35),

$$Y_{(m-1)ij}^2 = \sum_k W_{ki}^2 W_{kj}^2 + \sum_{k\ne \ell} W_{ki}W_{kj}W_{\ell i}W_{\ell j},$$

where $k$ and $\ell$ range in $\{1,2,...,m-1\}\setminus\{i,j\}$. Plugging it into (36), we have a decomposition

$$\sum_{m=1}^n \mathbb{E}(X_{n,m}^2|\mathcal{F}_{n,m-1}) = I_a + I_b, \tag{40}$$

where

$$I_a = \frac{1}{M_n}\sum_{m=1}^n \sum_{i<j\le m-1}\sum_{\substack{k\le m-1\\k\notin\{i,j\}}} W_{ki}^2 W_{kj}^2 \Omega_{mi}^*\Omega_{mj}^*,$$

$$I_b = \frac{1}{M_n}\sum_{m=1}^n \sum_{i<j\le m-1}\sum_{\substack{k,\ell\le m-1\\k,\ell\notin\{i,j\}}} W_{ki}W_{kj}W_{\ell i}W_{\ell j}\Omega_{mi}^*\Omega_{mj}^*.$$

Then,

$$\mathrm{Var}\Big(\sum_{m=1}^n \mathbb{E}(X_{n,m}^2|\mathcal{F}_{n,m-1})\Big)$$
$$= \mathrm{Var}(I_a) + \mathrm{Var}(I_b) + 2\mathrm{Cov}(I_a, I_b)$$
$$\le \Big(\sqrt{\mathrm{Var}(I_a)}+\sqrt{\mathrm{Var}(I_b)}\Big)^2 \tag{41}$$

It suffices to show that both $\mathrm{Var}(I_a) \to 0$ and $\mathrm{Var}(I_b) \to 0$.

Consider the variance of $I_a$. In the sum of $I_a$, all 4-cycles $(k,i,m,j)$ involved are selected in this way: We first select $m$, then select a pair $(i,j)$ from $\{1,2,\ldots,m-1\}$ and connect both $i$ and $j$ to $m$, and finally select $k$ to close the cycle. In fact, these 4-cycles can be selected in an alternative way: First, select a V-shape $(i,k,j)$ with $k$ being the middle point. Second, select $m > \max\{i,k,j\}$ to make the V-shape a cycle. Hence, we can rewrite

$$I_a = \frac{1}{M_n}\sum_{k=1}^n \sum_{\substack{1\le i<j\le n\\i\ne k, j\ne k}} W_{ki}^2 W_{kj}^2 \underbrace{\sum_{m>\max\{i,j,k\}} \Omega_{mi}^*\Omega_{mj}^*}_{\equiv b_{kij}}$$

The terms $W_{ki}^2 W_{kj}^2$ corresponding to different $k$ are independent of each other. We now fix $k$ and calculate the covariance between $W_{ki}^2 W_{kj}^2$ and $W_{ki'}^2 W_{kj'}^2$, for $(i,j) \ne (i',j')$. There are three cases. Case (i): $(i,j)=(i',j')$. In this case, $\mathrm{Var}(W_{ki}^2 W_{kj}^2) \le \mathbb{E}[W_{ki}^4 W_{kj}^4] \le \mathbb{E}[W_{ki}^2 W_{kj}^2] \le \Omega_{ki}^*\Omega_{kj}^*$. Case (ii): $i=i'$ but $j\ne j'$. In this case, we have $\mathrm{Cov}(W_{ki}^2 W_{kj}^2, W_{ki}^2 W_{kj'}^2) = \mathrm{Var}(W_{ki}^2)\cdot \mathbb{E}[W_{kj}^2]\mathbb{E}[W_{kj'}^2] \le \Omega_{ki}^*\Omega_{kj}^*\Omega_{kj'}^*$. Case (iii): $(i,j)\cap(i',j')=\emptyset$. The two terms are independent, and their covariance is zero. Combining the above gives

$$\mathrm{Var}(I_a) \le \frac{1}{M_n^2}\sum_{k=1}^n\Big(\sum_{\substack{1\le i<j\le n\\i\ne k, j\ne k}} b_{kij}^2 \Omega_{ki}^*\Omega_{kj}^*$$
$$+ \sum_{\substack{i,j,j'\in\{1,\ldots,n\}\setminus\{k\}\\i,j,j'\text{ are distinct}}} b_{kij}b_{kij'}\Omega_{ki}^*\Omega_{kj}^*\Omega_{kj'}^*\Big).$$

We now bound the right hand side. By condition (9), $\Omega_{ij}^* \le C\theta_i\theta_j$. Hence, $b_{kij} \le C\sum_{m>k}\theta_m^2\theta_i\theta_j \le C\|\theta\|^2\theta_i\theta_j$. As a result,

$$\mathrm{Var}(I_a) \le \frac{C}{M_n^2}\Big[\sum_{k,i,j}\|\theta\|^4\theta_k^2\theta_i^3\theta_j^3 + \sum_{k,i,j,j'}\|\theta\|^4\theta_k^3\theta_i^3\theta_j^2\theta_{j'}^2\Big]$$
$$\le \frac{C}{M_n^2}(\|\theta\|^6\|\theta\|_3^6 + \|\theta\|^8\|\theta\|_3^6).$$

By (7), $\|\theta\|\to\infty$, so the second term dominates. Moreover, since $\Omega_{ij}^* = \Omega_{ij}(1-\Omega_{ij}) \ge c\Omega_{ij}$ (in our setting, all $\Omega_{ij}$'s are bounded away from 1). As a result, we have $M_n \ge c\sum_{CC(I_n)}\Omega_{i_1i_2}\Omega_{i_2i_3}\Omega_{i_3i_4}\Omega_{i_4i_1} \ge C^{-1}n^4C_4$. By Lemma B.1, $n^4C_4 \asymp \|\theta\|^8$. Combining the above gives $\mathrm{Var} = O(\|\theta\|_3^6/\|\theta\|^8)$, i.e.,

$$\sqrt{\mathrm{Var}(I_a)} \le \frac{C\sum_i \theta_i^3}{(\sum_i \theta_i^2)^2} \le \frac{C\theta_{\max}}{\sum_i \theta_i^2} = o(1). \tag{42}$$

Consider the variance of $I_b$. Rewrite

$$I_b = \frac{1}{M_n}\sum_{k,\ell,i,j\text{ are distinct}} c_{k\ell ij}G_{k\ell ij},$$

where

$$G_{k\ell ij} \equiv W_{ki}W_{kj}W_{\ell i}W_{\ell j}, \quad c_{k\ell ij} = \sum_{\substack{m> \\ \max\{k,\ell,i,j\}}} \Omega^*_{mi}\Omega^*_{mj}.$$

Since $I_b$ has a mean zero, $\mathrm{Var}(I_b) = \mathbb{E}(I_b^2)$. Additionally, for 2 cycles $(k,\ell,i,j)$ and $(k',\ell',i',j')$, only when they are exactly equal, we have $\mathbb{E}[G_{k\ell ij}G_{k'\ell'i'j'}] \neq 0$. As a result,

$$\begin{aligned}\mathrm{Var}(I_b) &= \frac{1}{M_n} \sum_{\substack{k,\ell,i,j \text{ are distinct}}} c^2_{k\ell ij}\mathbb{E}[G^2_{k\ell ij}] \\ &= \frac{1}{M_n} \sum_{\substack{k,\ell,i,j \text{ are distinct}}} c^2_{k\ell ij}\Omega^*_{ki}\Omega^*_{kj}\Omega^*_{\ell i}\Omega^*_{\ell j}.\end{aligned}$$

Similarly to how we get the bound for $b_{kij}$, we can derive that $c_{k\ell ij} \leq C\|\theta\|^2\theta_i\theta_j$. Moreover, $\Omega^*_{ki}\Omega^*_{kj}\Omega^*_{\ell i}\Omega^*_{\ell j} \leq C\theta_i^2\theta_j^2\theta_k^2\theta_\ell^2$. Hence,

$$\mathrm{Var}(I_b) \leq \frac{C}{\|\theta\|^{16}} \sum_{k,\ell,i,j} \|\theta\|^4 \theta_k^2 \theta_\ell^2 \theta_i^4 \theta_j^4 \leq \frac{C\|\theta\|_4^8}{\|\theta\|^8}.$$

As a result,

$$\sqrt{\mathrm{Var}(I_b)} \leq \frac{C\sum_i \theta_i^4}{(\sum_i \theta_i^2)^2} \leq \frac{C\theta_{\max}^2}{\sum_i \theta_i^2} = o(1). \quad (43)$$

Plugging (42)-(43) into (41) gives (38). Combining (37) and (38), we have proved (a).

We now check (b). By the Cauchy-Schwarz inequality and the Chebyshev's inequality,

$$\begin{aligned}&\sum_{m=1}^n \mathbb{E}(X^2_{n,m}\mathbf{1}_{\{|X_{n,m}|>\epsilon\}}|\mathcal{F}_{n,m-1}) \\ &\leq \sum_{m=1}^n \sqrt{\mathbb{E}(X^4_{n,m}|\mathcal{F}_{n,m-1})}\sqrt{\mathbb{P}(|X_{n,m}|\geq \epsilon|\mathcal{F}_{n,m-1})} \\ &\leq \epsilon^{-2} \sum_{m=1}^n \mathbb{E}(X^4_{n,m}|\mathcal{F}_{n,m-1}).\end{aligned}$$

Therefore, it suffices to show that the right hand side converges to zero in probability. Then, it suffices to show that its $L^1$-norm converges to zero. Since the right hand is a nonnegative random variable, we only need to prove that its expectation converges to zero, i.e.,

$$\mathbb{E}\Big[\sum_{m=1}^n X^4_{n,m}\Big] = o(1). \quad (44)$$

We now prove (44). We use the expression of $X_{n,m}$ in (34). Conditioning on $\mathcal{F}_{n,m-1}$, the $Y_{(m-1)ij}$'s are non-random. It follows that

$$\mathbb{E}[X^4_{n,m}|\mathcal{F}_{n,m-1}] = \frac{1}{M_n^4} \sum_{\substack{i,j=1 \\ i\neq j}}^{m-1} Y^2_{(m-1)ij}\mathbb{E}[W^4_{mi}W^4_{mj}]$$

$$+ \frac{1}{M_n^4} \sum_{i=1}^{m-1} \sum_{\substack{j,j'=1 \\ j,j'\notin\{i\}}}^{m-1} Y_{(m-1)ij}Y_{(m-1)ij'}\mathbb{E}[W^4_{mi}W^2_{mj}W^2_{mj'}]$$

$$+ \frac{1}{M_n^4} \sum_{\substack{i,j,i',j'=1 \\ \text{distinct}}}^{m-1} Y_{(m-1)ij}Y_{(m-1)i'j'}\mathbb{E}[W^2_{mi}W^2_{mj}W^2_{mi'}W^2_{mj'}].$$

First, we shall use the independence across entries of $W$ and the fact that $\mathbb{E}[W^4_{ij}] \leq \mathbb{E}[W^2_{ij}] \leq \Omega_{ij} \leq C\theta_i\theta_j$. Second, in proving (39), we have seen that $\mathbb{E}[Y^2_{(m-1)ij}] = \sum_{k<m,k\notin\{i,j\}} \Omega^*_{ki}\Omega^*_{kj} \leq C\sum_k \theta_k^2\theta_i\theta_j \leq C\|\theta\|^2\theta_i\theta_j$. Third, from (35), it is easy to see that when $(i,j,i',j')$ are distinct, $\mathbb{E}[Y_{(m-1)ij}Y_{(m-1)i'j'}] = 0$; moreover, for $j \neq j'$, $\mathbb{E}[Y_{(m-1)ij}Y_{(m-1)ij'}] = \sum_k \mathbb{E}[W^2_{ki}]\mathbb{E}[W_{kj}W_{kj'}] = 0$. Last, in proving (42), we have seen that $M_n \geq c\|\theta\|^8$. Combining the above, we find that

$$\begin{aligned}\mathbb{E}[X^4_{n,m}] &= \frac{1}{M_n^2} \sum_{\substack{i,j=1 \\ i\neq j}}^{m-1} \mathbb{E}[Y^2_{(m-1)ij}]\mathbb{E}[W^4_{mi}W^4_{mj}] \\ &\leq \frac{C}{\|\theta\|^{16}} \sum_{i,j=1}^{m-1}(\|\theta\|^2\theta_i\theta_j)(\theta_m\theta_i)(\theta_m\theta_j) \\ &\leq C\theta_m^2/\|\theta\|^{10}.\end{aligned}$$

As a result,

$$\sum_{n=1}^n \mathbb{E}[X^4_{n,m}] \leq C\|\theta\|^{-8} = o(1).$$

This gives (44) and (b) follows. $\square$

## C. Proof of Theorem 3.3 and Corollary 3.1

Consider Theorem 3.3 first. For short, let

$$Z_n^{(m)} = \sqrt{\frac{B_{n,m}}{2m}}\widehat{C}_m^{-1/2}\widehat{\chi}_{gc}^{(m)}, \quad x_0^* = \mathbb{P}(Z_n^{(m)} \geq z_\alpha).$$

It suffices to show that under the null and alternative,

$$|x_0^* - \Phi(\delta_{gc}^{(m)} - z_\alpha)| \leq o(1). \quad (45)$$

Denote $a_n = (C_m/\widehat{C}_m)^{1/2}$ for short. It is seen that

$$a_n \xrightarrow{p} 1, \quad (46)$$

and

$$\frac{1}{a_n}Z_n^{(m)} = \sqrt{\frac{B_{n,m}}{2m}}C_m^{-1/2}\widehat{\chi}_{gc}^{(m)}. \quad (47)$$

Combining Theorem 3.1 and the proof of Theorem 3.2, we have shown that

$$\sqrt{\frac{B_{n,m}}{2m}}C_m^{-1/2}\big[\widehat{\chi}_{gc}^{(m)} - \chi_{gc,0}^{(m)}\big] \xrightarrow{p} N(0,1), \quad (48)$$

where by definitions,

$$\sqrt{\frac{B_{n,m}}{2m}} C_m^{-1/2} \chi_{gc,0}^{(m)} = \delta_{gc}^{(m)}. \tag{49}$$

Combining (47)-(49) gives

$$\frac{1}{a_n} Z_n^{(m)} - \delta_{gc}^{(m)} \xrightarrow{d} N(0,1). \tag{50}$$

Denote the CDF of $\frac{1}{a_n} Z_n^{(m)} - \delta_{gc}^{(m)}$ by $F_n$. Recall that $\Phi$ denotes the CDF of $N(0,1)$. It follows from (50) that

$$\sup_x |F_n(x) - \Phi(x)| \to 0. \tag{51}$$

We now rewrite

$$x_0^* = \mathbb{P}\left(\frac{1}{a_n} Z_n^{(m)} - \delta_{gc}^{(m)} \geq \frac{1}{a_n} z_\alpha - \delta_{gc}^{(m)}\right),$$

and introduce a proxy by

$$x_0 = \mathbb{P}\left(\frac{1}{a_n} Z_n^{(m)} - \delta_{gc}^{(m)} \geq z_\alpha - \delta_{gc}^{(m)}\right).$$

By triangle inequality,

$$|x_0^* - \Phi(\delta_{gc}^{(m)} - z_\alpha)| \leq |x_0^* - x_0| + |x_0 - \Phi(\delta_{gc}^{(m)} - z_\alpha)|. \tag{52}$$

where by (51),

$$|x_0 - \Phi(\delta_{gc}^{(m)} - z_\alpha)| \to 0. \tag{53}$$

Moreover, for any fixed $\epsilon > 0$, it is seen that

$$|x_0^* - x_0| \leq I + II,$$

where

$$I = \mathbb{P}(|a_n - 1| \geq \epsilon),$$

and

$II = \mathbb{P}(\frac{1}{a_n} Z_n^{(m)} - \delta_{gc}^{(m)}$ falls between $(1 \pm \epsilon)z_\alpha - \delta_{gc}^{(m)})$,

which by (51) does not exceed

$\mathbb{P}(N(0,1)$ falls between $(1 \pm \epsilon)z_\alpha - \delta_{gc}^{(m)}) + o(1)$;

note the first term does not exceed $(2/\sqrt{2\pi})z_\alpha \epsilon$. Combining these gives that for any $\epsilon > 0$,

$$|x_0^* - \Phi(\delta_{gc}^{(m)} - z_\alpha)| \leq (2/\sqrt{2\pi})z_\alpha \epsilon + \mathbb{P}(|a_n - 1| \geq \epsilon) + o(1).$$

Recall that $a_n \xrightarrow{p} 1$, the claim follows.

Next, consider Corollary 3.1. It is seen that $\delta_{gc}^{(m)} = 0$ under the null and that under the alternative,

$$\delta_{gc}^{(m)} \geq \sum_{k=2}^{K} \lambda_k^m / [\sum_{k=1}^{K} \lambda_k^m]^{1/2}.$$

When $m = 4$, by Lemma 6.1, $\delta_{gc}^{(4)} \geq c_4 \|\theta\|^4$ for some constant $c_4 > 0$. When $m = 3$ and $P$ is positive definite, $\lambda_k$ are the eigenvalues of $\Theta\Pi P\Pi'\Theta$, so for $1 \leq k \leq K$, $\lambda_k \geq 0$. Using Lemma 6.1, $\delta_{gc}^{(3)} \geq c_3 \|\theta\|^3$ for some constant $c_3$. Combining these with Theorem 3.3 gives the claim.

## D. Proof of Secondary Lemmas

### D.1. Proof of Lemma 6.1

We first consider the claim about $\lambda_k$'s. Recall that $\lambda_k$'s are the eigenvalues of the matrix $G^{1/2}PG^{1/2}$. First, we have $\|G\| \leq \sum_{k,\ell} G(k,\ell) = \sum_{k,\ell} \sum_i \theta_i^2 \pi_i(k) \pi_i(\ell) = \sum_i \theta_i^2 \sum_{k,\ell} \pi_i(k) \pi_i(\ell) = \|\theta\|^2$. Second, let $g_k = \sum_{i \in \mathcal{N}_k} \theta_i^2$ for $1 \leq k \leq K$, and write $\Theta = \Theta_1 + \Theta_2$, where $\Theta_1(i,i) = \theta_i \cdot 1\{i \in \cup_{k=1}^{K} \mathcal{N}_k\}$ and $\Theta_2 \equiv \Theta - \Theta_1$. It yields that $G = \Pi'\Theta_1^2\Pi + \Pi'\Theta_2^2\Pi = \text{diag}(g_1, \cdots, g_K) + \Pi'\Theta_2^2\Pi$. Hence, $\lambda_{\min}(G) \geq \min_{1 \leq k \leq K} g_k \geq c_2 \|\theta\|^2$, where the last inequality is from condition (8). Combining the above gives

$$c_2 \|\theta\|^2 \leq \lambda_{\min}(G) \leq \|\theta\|^2. \tag{54}$$

Using condition (9), we find that $|\lambda_k| \leq \|PG\| \leq C\|G\| = O(\|\theta\|^2)$. Additionally, since $|\lambda_k|^2$ is an eigenvalue of $(G^{1/2}PG^{1/2})^2 = G^{1/2}PGPG^{1/2}$, we then have $|\lambda_k|^2 \geq \lambda_{\min}(G) \cdot \lambda_{\min}(PGP) \geq \lambda_{\min}^2(G) \cdot s_{\min}^2(P) \geq c_1^2 c_2^2 \|\theta\|^4$. It gives

$$C^{-1}\|\theta\|^2 \leq |\lambda_k| \leq C\|\theta\|^2, \qquad 1 \leq k \leq K.$$

We then consider the claim about $\eta$. Since $\max_k |\eta'\xi_k|^2$ is upper bounded by $\sum_k |\eta'\xi_k|^2$ and lower bounded by $K^{-1} \sum_k |\eta'\xi_k|^2$, it suffices to show that

$$C^{-1}\|\theta\|_1^2 \leq \sum_{1 \leq k \leq K} |\eta'\xi_k|^2 \leq C\|\theta\|_1^2. \tag{55}$$

Since $\xi_1, \ldots, \xi_K$ form an orthonormal basis,

$$\sum_{1 \leq k \leq K} |\eta'\xi_k|^2 = \|\eta\|^2 = 1_n'\Theta\Pi G^{-1}\Pi'\Theta 1_n.$$

It follows from (54) that the right hand side has the same order as $\|\theta\|^{-2}\|\Pi'\Theta 1_n\|^2$. Write $v = \Pi'\Theta 1_n$. For $1 \leq k \leq K$, $v(k) = \sum_i \pi_i(k)\theta_i$. It follows that $v(k) \leq \|\theta\|_1$. At the same time, $\sum_{k=1}^{K} v^2(k) \geq \frac{\left(\sum_{k=1}^{K} v(k)\right)^2}{K} = \frac{\|\theta\|_1^2}{K}$, where we've used Cauchy-Schwarz inequality.

It follows that

$$C^{-1}\|\theta\|_1^2 \leq \|\Pi'\Theta 1_n\|^2 \leq C\|\theta\|_1^2.$$

Hence, (55) follows. $\square$

### D.2. Proof of Lemma B.1

Consider the first item. By (16) of (Jin et al., 2018b),

$$C_4 = \frac{1}{B_{n,4}}\left[\sum_{k=1}^{K} \lambda_k^4 + O(\|\theta\|_4^4 \|\theta\|^4)\right],$$

where we note $B_{n,4} \sim n^{-4}$. First, by Lemma 6.1 of (Jin et al., 2018b),

$$\sum_{k=1}^{K} \lambda_k^4 \asymp \|\theta\|^8,$$

Second, by (7) of (Jin et al., 2018b), $\theta_{max} \leq \|\theta\|_3 \to 0$, so it is seem $\|\theta\|_4^4 \leq \theta_{max}\|\theta\|_3^3 \leq o(1)$, and so $\|\theta\|_4^4\|\theta\|^4 \leq o(\|\theta\|^4)$. Combining these give the claim.

Consider the second item. By (17) of (Jin et al., 2018b),

$$L_2 = \frac{1}{B_{n,3}}\left[\sum_{k=1}^K \lambda_k^2(\eta,\xi_k)^2 + O(\|\theta\|_1^2\|\theta\|_4^4\|\theta\|^{-2})\right],$$

where by Lemma 6.1 of (Jin et al., 2018b),

$$\sum_{k=1}^K \lambda_k^2(\eta,\xi_k)^2 \asymp \|\theta\|^2\|\theta\|_1^2.$$

By similar argument, $\|\theta\|_1^2\|\theta\|_4^4\|\theta\|^{-2} \leq o(\|\theta\|^2\|\theta\|_1^2)$, so the claim follows by noting $B_{n,3} \sim n^{-3}$.

Consider the third item. By similar argument, it is seen that

$$L_3 = \frac{1}{B_{n,4}}\left[\sum_{k=1}^K \lambda_k^3(\eta,\xi_k)^2 + O(\|\theta\|_1^2\|\theta\|_4^4)\right]$$
$$\leq Cn^{-4}\|\theta\|_1^2\|\theta\|^4.$$

For the lower bound, we use a different proof as $\lambda_k$ may be negative. By $L_3 = E[\hat{L}_3]$ and $E[A_{ij}] = \Omega_{ij}$ when $i \neq j$,

$$L_3 = \frac{1}{B_{n,4}}\sum_{\substack{1\leq i_1,i_2,i_3,i_4\leq n \\ \text{are distinct}}} \Omega_{i_1i_2}\Omega_{i_2i_3}\Omega_{i_3i_4}.$$

As before, let $\mathcal{N}_1$ denote the set of pure nodes in community 1. It is not hard to see that

$$L_3 \geq \sum_{k=1}^K \sum_{\substack{i_1,i_2,i_3,i_4\in\mathcal{N}_k \\ \text{are distinct}}} \Omega_{i_1i_2}\Omega_{i_2i_3}\Omega_{i_3i_4}.$$

In our model, all diagonal entries of $P$ are 1, so for any $i,j \in \mathcal{N}_1$, $\Omega_{ij} = \theta_i\theta_j$. Therefore,

$$L_3 \geq \sum_{k=1}^K \sum_{\substack{i_1,i_2,i_3,i_4\in\mathcal{N}_k \\ \text{are distinct}}} \theta_{i_1}\theta_{i_4}\theta_{i_2}^2\theta_{i_3}^2. \quad (56)$$

Now, we can lower bound the right hand side of (56) by

$$I - II - III - IV,$$

where

$$I = \sum_{k=1}^K \sum_{i_1,i_2,i_3,i_4\in\mathcal{N}_k} \theta_{i_1}\theta_{i_4}\theta_{i_2}^2\theta_{i_3}^2,$$

$$II = \sum_{k=1}^K \sum_{\substack{i_1,i_2,i_3,i_4\in\mathcal{N}_k \\ i_1=i_4}} \theta_{i_1}\theta_{i_4}\theta_{i_2}^2\theta_{i_3}^2,$$

$$III = \sum_{\substack{i_1,i_2,i_3,i_4\in\mathcal{N}_1 \\ i_2=i_3}} \theta_{i_1}\theta_{i_4}\theta_{i_2}^2\theta_{i_3}^2,$$

and

$$IV = 4\sum_{k=1}^K \sum_{\substack{i_1,i_2,i_3,i_4\in\mathcal{N}_k \\ i_1=i_2}} \theta_{i_1}\theta_{i_4}\theta_{i_2}^2\theta_{i_3}^2.$$

First, by (8) of (Jin et al., 2018b),

$$I = \sum_{k=1}^K \Big(\sum_{i\in\mathcal{N}_k}\theta_i\Big)^2\Big(\sum_{i\in\mathcal{N}_k}\theta_i^2\Big)^2$$
$$\geq C\|\theta\|^4\sum_{k=1}^K \Big(\sum_{i\in\mathcal{N}_k}\theta_i\Big)^2$$
$$\geq C\|\theta\|_1^2\|\theta\|^4,$$

where the last inequality is Cauchy-Schwarz inequality.

Second, by (7) of (Jin et al., 2018b) that $\theta_{\max} \leq \|\theta\|_3 = o(1)$, we obtain $\|\theta\|^2 \leq o(1) \cdot \|\theta\|_1$ (note $\|\theta\| \to \infty$), and

$$II \leq C\sum_{k=1}^K \sum_{i_1,i_2,i_3\in\mathcal{N}_k} \theta_{i_1}\theta_{i_2}^2\theta_{i_3}^2$$
$$\leq C\|\theta\|_1\|\theta\|^4$$
$$= o(\|\theta\|_1^2\|\theta\|^4).$$

Similarly, we have $\|\theta\|_3^3 \leq o(1)\cdot\|\theta\|_2^2$ and $\|\theta\|_4^4 \leq o(1)\cdot\|\theta\|_2^2$, which implies

$$III = o(\|\theta\|_1^2\|\theta\|^4), \quad \text{and} \quad IV = o(\|\theta\|_1^2\|\theta\|^4).$$

Combining these gives $L_3 \geq c\|\theta\|_1^2\|\theta\|^4$, and the claim follows.

We now prove the next three items (on the variances). In the Proof of Lemma B.2, we've already shown that $\frac{1}{B_{n,4}}\sum_{\substack{i_1,\cdots,i_4 \\ \text{distinct}}} G_{i_1i_2i_3i_4}(W)$ is the dominating term of $(\hat{C}_4 - C_4)$, and that

$$\text{Var}\Big(\frac{1}{B_{n,4}}\sum_{\substack{i_1,\cdots,i_4 \\ \text{distinct}}} G_{i_1i_2i_3i_4}(W)\Big)$$
$$\leq Cn^{-4}\sum_{\substack{i_1,\cdots,i_4 \\ \text{distinct}}} G_{i_1i_2i_3i_4}(\Omega)$$
$$\leq Cn^{-4}\sum_{i_1,\cdots,i_4}\theta_{i_1}^2\theta_{i_2}^2\theta_{i_3}^2\theta_{i_4}^2 = Cn^{-4}\|\theta\|^8.$$

Combining it with $C_4 \asymp n^{-4}\|\theta\|^8$, we get $\text{Var}(\hat{C}_4) = O(n^{-4}C_4)$.

Consider $\text{Var}(\hat{L}_2)$. By definitions and that $B_{n,m} \asymp n^m$, we bound

$$\mathbb{E}(\hat{L}_2-L_2)^2 \leq Cn^{-6}\mathbb{E}[\sum_{i_1<i_2<i_3}(A_{i_1i_2}A_{i_2i_3}-\Omega_{i_1i_2}\Omega_{i_2i_3})]^2.$$
(57)

Recall that when $i \neq j$, $A_{ij} = \Omega_{ij} + W_{ij}$. Since for any numbers $x, y, a, b$, $(a + x)(b + y) - ab = xy + ay + bx$, we can write

$$\sum_{i_1 < i_2 < i_3} (A_{i_1 i_2} A_{i_2 i_3} - \Omega_{i_1 i_2} \Omega_{i_2 i_3}) = I + II + III,$$

where

$$I = \sum_{i_1 < i_2 < i_3} W_{i_1 i_2} W_{i_2 i_3},$$

$$II = \sum_{i_1 < i_2 < i_3} \Omega_{i_1 i_2} W_{i_2 i_3},$$

and

$$III = \sum_{i_1 < i_2 < i_3} \Omega_{i_2 i_3} W_{i_1 i_2}.$$

Inserting this into (57) and using Cauchy-Schwarz inequality,

$$\mathbb{E}(\widehat{L}_2 - L_2)^2 \leq Cn^{-6}(\mathbb{E}[(I)^2] + \mathbb{E}[(II)^2] + \mathbb{E}[(III)^2]).$$

It then suffices to show

$$\mathbb{E}[(I)^2] \lesssim \|\theta\|_1^3 \|\theta\|_3^3, \quad (58)$$

$$\mathbb{E}[(II)^2] \lesssim \|\theta\|_1^3 \|\theta\|_3^3, \quad (59)$$

and

$$\mathbb{E}[(III)^2] \lesssim \|\theta\|_1^3 \|\theta\|_3^3. \quad (60)$$

We now show (58)-(60) separately.

Consider (58). Note that for two sets of indices $(i_1, i_2, i_3)$ and $(j_1, j_2, j_3)$ such that $i_1 < i_2 < i_3$, $j_1 < j_2 < j_3$, by basic statistics, we have that when $(i_1, i_2, i_3) \neq (j_1, j_2, j_3)$,

$$\mathbb{E}[W_{i_1 i_2} W_{i_2 i_3} W_{j_1 j_2} W_{j_2 j_3}] = 0.$$

and when $(i_1, i_2, i_3) = (j_1, j_2, j_3)$,

$$\mathbb{E}[W_{i_1 i_2} W_{i_2 i_3} W_{j_1 j_2} W_{j_2 j_3}] = E[W_{i_1 i_2}^2 W_{i_2 i_3}^2]$$
$$= \Omega_{i_1 i_2}(1 - \Omega_{i_1 i_2}) \Omega_{i_2 i_3}(1 - \Omega_{i_2 i_3}).$$

Therefore,

$$\mathbb{E}[(I)]^2 = \sum_{i_1 < i_2 < i_3} \sum_{j_1 < j_2 < j_3} E[W_{i_1 i_2} W_{i_2 i_3} W_{j_1 j_2} W_{j_2 j_3}]$$
$$\leq \sum_{i_1 < i_2 < i_3} \mathbb{E}[W_{i_1 i_2}^2 W_{i_2 i_3}^2]$$
$$\leq \sum_{i_1 < i_2 < i_3} \Omega_{i_1 i_2}(1 - \Omega_{i_1 i_2}) \Omega_{i_2 i_3}(1 - \Omega_{i_2 i_3}).$$

Recall that for any $i < j$,

$$\Omega_{ij}(1 - \Omega_{ij}) \leq \Omega_{ij} \leq \theta_i \theta_j,$$

it follows that

$$\mathbb{E}[(I)]^2 \leq \sum_{i_1 < i_2 < i_3} \theta_{i_1} \theta_{i_2} \theta_{i_2} \theta_{i_3} \leq \|\theta\|_1^2 \|\theta\|_2^2,$$

and the claim follows by Cauchy Schwarz inequality that $\|\theta\|_1^2 \|\theta\|_2^2 \leq \|\theta\|_1^2 (\|\theta\|_1 \|\theta\|_3^3) = \|\theta\|_1^3 \|\theta\|_3^3$.

Consider (59)-(60). Since the proofs are similar, we only show (59).

Note that for two sets of indices $(i_1, i_2, i_3)$ and $(j_1, j_2, j_3)$ such that $i_1 < i_2 < i_3$, $j_1 < j_2 < j_3$, by basic statistics, we have that when $(i_2, i_3) \neq (j_2, j_3)$,

$$\mathbb{E}[W_{i_2 i_3} W_{j_2 j_3}] = 0.$$

and when $(i_2, i_3) = (j_2, j_3)$,

$$\mathbb{E}[W_{i_2 i_3} W_{j_2 j_3}] = \mathbb{E}[W_{i_2 i_3}^2] = \Omega_{i_2 i_3}(1 - \Omega_{i_2 i_3}).$$

Therefore,

$$\mathbb{E}[(II)^2] = \sum_{i_1 < i_2 < i_3} \sum_{j_1 < j_2 < j_3} \mathbb{E}[\Omega_{i_1 i_2} \Omega_{j_1 j_2} W_{i_2 i_3} W_{j_2 j_3}]$$
$$= \sum_{i_1 < i_2 < i_3} \mathbb{E}[\Omega_{i_1 i_2} W_{i_2 i_3} \sum_{j_1 < j_2 < j_3} (\Omega_{j_1 j_2} W_{j_2 j_3})]$$
$$= \sum_{i_1 < i_2 < i_3} \mathbb{E}[\Omega_{i_1 i_2} W_{i_2 i_3}^2 (\sum_{j_1 < i_2} \Omega_{j_1 i_2})]$$
$$= \sum_{i_1 < i_2 < i_3} \mathbb{E}[\Omega_{i_1 i_2} \Omega_{i_2 i_3}(1 - \Omega_{i_2 i_3})(\sum_{j_1 < i_2} \Omega_{j_1 i_2})]$$

Again by $\Omega_{ij}(1 - \Omega_{ij}) \leq \Omega_{ij} \leq \theta_i \theta_j$ for any $i < j$, we find

$$\mathbb{E}[(II)^2] \leq \sum_{i_1 < i_2 < i_3} \mathbb{E}[\Omega_{i_1 i_2} \Omega_{i_2 i_3}(\sum_{j_1 < i_2} \Omega_{j_1 i_2})]$$
$$\leq \sum_{i_1 < i_2 < i_3} \theta_{i_1} \theta_{i_2}^2 \theta_{i_3}(\sum_{j_1 < i_2} \theta_{j_1} \theta_{i_2})$$
$$\leq \|\theta\|_1^3 \|\theta\|_3^3.$$

Last, we prove the claim on $\text{Var}(\widehat{L}_3)$. It suffices to control the covariance between $(A_{i_1 i_2} A_{i_2 i_3} A_{i_3 i_4})$ and $(A_{j_1 j_2} A_{j_2 j_3} A_{j_3 j_4})$. To be more specific, define the set

$$\mathcal{J} = \{(i_1, i_2), (i_2, i_3), (i_3, i_4), (j_1, j_2), (j_2, j_3), (j_3, j_4)\},$$

whose elements are pairs of unordered integers, i.e. we treat $(i_1, i_2)$ and $(i_2, i_1)$ as the same element.

Let $|\mathcal{J}|$ be the number of distinct elements of $\mathcal{J}$, where $3 \leq |\mathcal{J}| \leq 6$ under the condition that $i_1 < i_2 < i_3 < i_4$ and $j_1 < j_2 < j_3 < j_4$. To control the variance of $\widehat{L}_3$, it suffices to bound the following quantity

$$\sum_{s=3}^{6} \sum_{|\mathcal{J}|=s} \text{Cov}(A_{i_1 i_2} A_{i_2 i_3} A_{i_3 i_4}, A_{j_1 j_2} A_{j_2 j_3} A_{j_3 j_4}).$$

Furthermore, it suffices to show for $3 \leq s \leq 6$,

$$\sum_{|\mathcal{J}|=s} \mathrm{Cov}(A_{i_1 i_2} A_{i_2 i_3} A_{i_3 i_4}, A_{j_1 j_2} A_{j_2 j_3} A_{j_3 j_4}) \lesssim \|\theta\|_1^4 \|\theta\|_3^6. \tag{61}$$

When $|\mathcal{J}| = 6$, it's not hard to see $(A_{i_1 i_2} A_{i_2 i_3} A_{i_3 i_4})$ and $(A_{j_1 j_2} A_{j_2 j_3} A_{j_3 j_4})$ are independent because the six elements in $\mathcal{J}$ are all distinct, which indicates

$$\sum_{|\mathcal{J}|=6} \mathrm{Cov}(A_{i_1 i_2} A_{i_2 i_3} A_{i_3 i_4}, A_{j_1 j_2} A_{j_2 j_3} A_{j_3 j_4}) = 0.$$

The following basic property is frequently used in the discussion of remaining cases. For non-negative random variables $X$ and $Y$, we have

$$\mathrm{Cov}(X, Y) \leq \mathbb{E}[XY]. \tag{62}$$

Consider the case where $|\mathcal{J}| = 5$. By symmetry, it's enough to consider three situations where $(i_1, i_2) = (j_1, j_2)$, $(i_1, i_2) = (j_2, j_3)$ and $(i_2, i_3) = (j_2, j_3)$, separately.

If $(i_1, i_2) = (j_1, j_2)$, we have

$$\sum_{(i_1,i_2)=(j_1,j_2)} \mathrm{Cov}(A_{i_1 i_2} A_{i_2 i_3} A_{i_3 i_4}, A_{j_1 j_2} A_{j_2 j_3} A_{j_3 j_4})$$
$$\leq \sum_{(i_1,i_2)=(j_1,j_2)} \mathbb{E}[A_{i_1 i_2} A_{i_2 i_3} A_{i_3 i_4} A_{j_2 j_3} A_{j_3 j_4}]$$
$$\leq C \sum_{(i_1,i_2)=(j_1,j_2)} \theta_{i_1} \theta_{i_2}^2 \theta_{i_3}^2 \theta_{i_4} \theta_{j_2} \theta_{j_3}^2 \theta_{j_4}$$
$$\leq C \sum \theta_{i_1} \theta_{i_2}^3 \theta_{i_3}^2 \theta_{i_4} \theta_{j_3}^2 \theta_{j_4}$$
$$= C \|\theta\|_1^3 \|\theta\|^4 \|\theta\|_3^3 \leq C \|\theta\|_1^4 \|\theta\|_3^6,$$

where the last inequality is due to $\|\theta\|^4 \leq \|\theta\|_1 \|\theta\|_3^3$ by Cauchy-Schwarz inequality.

If $(i_1, i_2) = (j_2, j_3)$, we have

$$\sum_{(i_1,i_2)=(j_2,j_3)} \mathrm{Cov}(A_{i_1 i_2} A_{i_2 i_3} A_{i_3 i_4}, A_{j_1 j_2} A_{j_2 j_3} A_{j_3 j_4})$$
$$\leq \sum_{(i_1,i_2)=(j_2,j_3)} \mathbb{E}[A_{i_1 i_2} A_{i_2 i_3} A_{i_3 i_4} A_{j_1 j_2} A_{j_3 j_4}]$$
$$\leq C \sum_{(i_1,i_2)=(j_2,j_3)} \theta_{i_1} \theta_{i_2}^2 \theta_{i_3}^2 \theta_{i_4} \theta_{j_1} \theta_{j_2} \theta_{j_3} \theta_{j_4}$$
$$= C \sum_{i_1,\cdots,i_4,j_1,j_4} \theta_{i_1}^2 \theta_{i_2}^3 \theta_{i_3}^2 \theta_{i_4} \theta_{j_1} \theta_{j_4}$$
$$= C \|\theta\|_1^3 \|\theta\|^4 \|\theta\|_3^3 \leq C \|\theta\|_1^4 \|\theta\|_3^6,$$

where the last inequality is due to $\|\theta\|^4 \leq \|\theta\|_1 \|\theta\|_3^3$.

If $(i_2, i_3) = (j_2, j_3)$, we have

$$\sum_{(i_2,i_3)=(j_2,j_3)} \mathrm{Cov}(A_{i_1 i_2} A_{i_2 i_3} A_{i_3 i_4}, A_{j_1 j_2} A_{j_2 j_3} A_{j_3 j_4})$$
$$\leq \sum_{(i_2,i_3)=(j_2,j_3)} \mathbb{E}[A_{i_1 i_2} A_{i_2 i_3} A_{i_3 i_4} A_{j_1 j_2} A_{j_3 j_4}]$$
$$\leq C \sum_{(i_2,i_3)=(j_2,j_3)} \theta_{i_1} \theta_{i_2}^2 \theta_{i_3}^2 \theta_{i_4} \theta_{j_1} \theta_{j_2} \theta_{j_3} \theta_{j_4}$$
$$= C \sum_{i_1,\cdots,i_4,j_1,j_4} \theta_{i_1} \theta_{i_2}^3 \theta_{i_3}^3 \theta_{i_4} \theta_{j_1} \theta_{j_4} = C \|\theta\|_1^4 \|\theta\|_3^6.$$

Combining above three inequalities, we derive

$$\sum_{|\mathcal{J}|=5} \mathrm{Cov}(A_{i_1 i_2} A_{i_2 i_3} A_{i_3 i_4}, A_{j_1 j_2} A_{j_2 j_3} A_{j_3 j_4}) \leq C \|\theta\|_1^4 \|\theta\|_3^6.$$

Consider the case where $|\mathcal{J}| = 4$. By symmetry, $\mathcal{J}$ either equals to $\mathcal{J}_1 = \{(i_1, i_2), (i_2, i_3), (i_3, i_4), (j_1, j_2)\}$ or $\mathcal{J}_2 = \{(i_1, i_2), (i_2, i_3), (i_3, i_4), (j_2, j_3)\}$.

Therefore, we decompose and bound

$$\sum_{|\mathcal{J}|=4} \mathrm{Cov}(A_{i_1 i_2} A_{i_2 i_3} A_{i_3 i_4}, A_{j_1 j_2} A_{j_2 j_3} A_{j_3 j_4})$$
$$\lesssim \sum_{\mathcal{J}_1} \mathrm{Cov}(A_{i_1 i_2} A_{i_2 i_3} A_{i_3 i_4}, A_{j_1 j_2} A_{j_2 j_3} A_{j_3 j_4})$$
$$+ \sum_{\mathcal{J}_2} \mathrm{Cov}(A_{i_1 i_2} A_{i_2 i_3} A_{i_3 i_4}, A_{j_1 j_2} A_{j_2 j_3} A_{j_3 j_4})$$
$$\leq \sum_{\mathcal{J}_1} \mathbb{E}[A_{i_1 i_2} A_{i_2 i_3} A_{i_3 i_4} A_{j_1 j_2}] + \sum_{\mathcal{J}_2} \mathbb{E}[A_{i_1 i_2} A_{i_2 i_3} A_{i_3 i_4} A_{j_2 j_3}]$$

It then suffices to show

$$\sum_{\mathcal{J}_1} \mathbb{E}[A_{i_1 i_2} A_{i_2 i_3} A_{i_3 i_4} A_{j_1 j_2}] \leq C \|\theta\|_1^4 \|\theta\|_3^6, \tag{63}$$

and

$$\sum_{\mathcal{J}_2} \mathbb{E}[A_{i_1 i_2} A_{i_2 i_3} A_{i_3 i_4} A_{j_2 j_3}] \leq C \|\theta\|_1^4 \|\theta\|_3^6, \tag{64}$$

For (63), $j_2$ must equal to one of $i_1, \cdots, i_4$ since $(j_2, j_3)$ equals to some $(i_s, i_{s+1})$ by definition of $\mathcal{J}_1$. By symmetry, we only need to consider $j_2 = i_1$ and $j_2 = i_2$. Again by $\Omega_{ij} \leq \theta_i \theta_j$, we obtain

$$\sum_{\mathcal{J}_1} \mathbb{E}[A_{i_1 i_2} A_{i_2 i_3} A_{i_3 i_4} A_{j_1 j_2}]$$
$$\leq \sum_{j_2=i_1} \theta_{i_1} \theta_{i_2}^2 \theta_{i_3}^2 \theta_{i_4} \theta_{j_1} \theta_{j_2} + \sum_{j_2=i_2} \theta_{i_1} \theta_{i_2}^2 \theta_{i_3}^2 \theta_{i_4} \theta_{j_1} \theta_{j_2}$$
$$\leq \sum \theta_{i_1}^2 \theta_{i_2}^2 \theta_{i_3}^2 \theta_{i_4} \theta_{j_1} + \sum \theta_{i_1} \theta_{i_2}^3 \theta_{i_3}^2 \theta_{i_4} \theta_{j_1}$$
$$= \|\theta\|_1^2 \|\theta\|^6 + C \|\theta\|_1^3 \|\theta\|^2 \|\theta\|_3^3$$
$$\leq \|\theta\|_1^4 \|\theta\|_3^6.$$

Here we explain the last inequality. By Cauchy-Schwartz inequality, $\|\theta\|^4 \leq \|\theta\|_1 \|\theta\|_3^3$. Combining with (7) that $\|\theta\| \to \infty$, $\|\theta\|_1^2 \|\theta\|^6 \lesssim \|\theta\|_1^2 \|\theta\|^8 \leq \|\theta\|_1^4 \|\theta\|^6$. Moreover, $\|\theta\|_1^3 \|\theta\|^2 \|\theta\|_3^3 \leq \|\theta\|_1^3 \|\theta\|_3^3 (\|\theta\|^4) \leq \|\theta\|_1^4 \|\theta\|_3^6$.

For (64), we similarly found $j_2$, $j_3$ must equal to some $i_1, \cdots, i_4$. By (7), $\theta_{j_3} \leq C$. Thus we only need to discuss the cases where $j_2 = i_1$ or $j_2 = i_2$.

$$\sum_{\mathcal{J}_2} \mathbb{E}\big[A_{i_1 i_2} A_{i_2 i_3} A_{i_3 i_4} A_{j_2 j_3}\big]$$
$$\leq \sum_{j_2 = i_1} \theta_{i_1} \theta_{i_2}^2 \theta_{i_3}^2 \theta_{i_4} \theta_{j_2} + \sum_{j_2 = i_2} \theta_{i_1} \theta_{i_2}^2 \theta_{i_3}^2 \theta_{i_4} \theta_{j_2}$$
$$\leq \sum \theta_{i_1}^2 \theta_{i_2}^2 \theta_{i_3}^2 \theta_{i_4} \theta_{j_2} + \sum \theta_{i_1} \theta_{i_2}^3 \theta_{i_3}^2 \theta_{i_4} \theta_{j_2}$$
$$= C\|\theta\|_1^2 \|\theta\|^6 + C\|\theta\|_1^3 \|\theta\|^2 \|\theta\|^3 \leq C\|\theta\|_1^4 \|\theta\|_3^6,$$

where the last inequality has been explained in the proof of (63).

Combining (63) and (64), we bound

$$\sum_{|\mathcal{J}|=4} \mathrm{Cov}(A_{i_1 i_2} A_{i_2 i_3} A_{i_3 i_4}, A_{j_1 j_2} A_{j_2 j_3} A_{j_3 j_4}) \leq C\|\theta\|_1^4 \|\theta\|_3^6.$$

Finally, consider the case where $|\mathcal{J}| = 3$. In this case, the covariance is in fact variance. Therefore,

$$\sum_{|\mathcal{J}|=3} \mathrm{Cov}(A_{i_1 i_2} A_{i_2 i_3} A_{i_3 i_4}, A_{j_1 j_2} A_{j_2 j_3} A_{j_3 j_4})$$
$$= \sum_{i_1, \cdots, i_4} \mathrm{Var}(A_{i_1 i_2} A_{i_2 i_3} A_{i_3 i_4})$$
$$\leq \sum_{i_1, \cdots, i_4} \mathbb{E}\big[A_{i_1 i_2} A_{i_2 i_3} A_{i_3 i_4}\big]$$
$$\leq \sum_{i_1, \cdots, i_4} \theta_{i_1} \theta_{i_2}^2 \theta_{i_3}^2 \theta_{i_4}$$
$$= C\|\theta\|_1^2 \|\theta\|^2 \lesssim C\|\theta\|_1^2 \|\theta\|^6 \leq C\|\theta\|_1^4 \|\theta\|_3^6,$$

where the second last inequality is by (7) that $\|\theta\| \to \infty$ and last inequality is Cauchy-Schwarz inequality.

This proves (61). $\square$

### D.3. Proof of Lemma B.2

Write for short $T_n = \frac{\sqrt{B_{n,4}}}{\sqrt{C_4}} (\widehat{C}_4 - C_4)$. We introduce some useful notations. For any $n \times n$ matrix $M$ and distinct indices $(i_1, i_2, i_3, i_4)$, define

$$G_{i_1 i_2 i_3 i_4}(M) = M_{i_1 i_2} M_{i_2 i_3} M_{i_3 i_4} M_{i_4 i_1},$$
$$G(M) = \sum_{(i_1, i_2, i_3, i_4) \in CC(I_n)} G_{i_1 i_2 i_3 i_4}(M).$$

Additionally, let $W = A - \Omega$ and let $\Omega^*$ be the matrix where $\Omega_{ij}^* = \Omega_{ij}(1 - \Omega_{ij})$ for all $1 \leq i, j \leq n$. We now rewrite

$$T_n = \frac{G(A) - G(\Omega)}{\sqrt{G(\Omega)}}, \qquad S_{n,n} = \frac{G(W)}{\sqrt{G(\Omega^*)}}. \tag{65}$$

Therefore,

$$T_n - S_{n,n} = \frac{G(A) - G(\Omega) - G(W)}{\sqrt{G(\Omega)}} + S_{n,n} \left[\frac{\sqrt{G(\Omega^*)}}{\sqrt{G(\Omega)}} - 1\right]$$
$$\equiv J_1 + S_{n,n} \cdot J_2.$$

In the proof of Theorem 3.2, we have shown $S_{n,n} \xrightarrow{d} N(0,1)$. Hence, to show $(T_n - S_{n,n}) \xrightarrow{p} 0$, it suffices to show that

$$J_1 \xrightarrow{p} 0 \tag{66}$$

and

$$J_2 \to 0. \tag{67}$$

First, we prove (66). We can decompose $G_{i_1 i_2 i_3 i_4}(A) - G_{i_1 i_2 i_3 i_4}(\Omega) - G_{i_1 i_2 i_3 i_4}(W)$ as the sum of three terms

$$\Delta_{i_1 i_2 i_3 i_4}^{(1)} = W_{i_1 i_2} \Omega_{i_2 i_3} \Omega_{i_3 i_4} \Omega_{i_4 i_1} + \Omega_{i_1 i_2} W_{i_2 i_3} \Omega_{i_3 i_4} \Omega_{i_4 i_1}$$
$$+ \Omega_{i_1 i_2} \Omega_{i_2 i_3} W_{i_3 i_4} \Omega_{i_4 i_1} + \Omega_{i_1 i_2} \Omega_{i_2 i_3} \Omega_{i_3 i_4} W_{i_4 i_1},$$
$$\Delta_{i_1 i_2 i_3 i_4}^{(2)} = W_{i_1 i_2} W_{i_2 i_3} \Omega_{i_3 i_4} \Omega_{i_4 i_1} + W_{i_1 i_2} \Omega_{i_2 i_3} W_{i_3 i_4} \Omega_{i_4 i_1}$$
$$+ W_{i_1 i_2} \Omega_{i_2 i_3} \Omega_{i_3 i_4} W_{i_4 i_1} + \Omega_{i_1 i_2} W_{i_2 i_3} W_{i_3 i_4} \Omega_{i_4 i_1}$$
$$+ \Omega_{i_1 i_2} W_{i_2 i_3} \Omega_{i_3 i_4} W_{i_4 i_1} + \Omega_{i_1 i_2} \Omega_{i_2 i_3} W_{i_3 i_4} W_{i_4 i_1},$$
$$\Delta_{i_1 i_2 i_3 i_4}^{(3)} = \Omega_{i_1 i_2} W_{i_2 i_3} W_{i_3 i_4} W_{i_4 i_1} + W_{i_1 i_2} \Omega_{i_2 i_3} W_{i_3 i_4} W_{i_4 i_1}$$
$$+ W_{i_1 i_2} W_{i_2 i_3} \Omega_{i_3 i_4} W_{i_4 i_1} + W_{i_1 i_2} W_{i_2 i_3} W_{i_3 i_4} \Omega_{i_4 i_1}.$$

It is easy to see that

$$\mathbb{E}\left[\sum_{CC(I_n)} \Delta_{i_1 i_2 i_3 i_4}^{(1)}\right] = 0. \tag{68}$$

We then study the variance of this term. Note that the four terms in $\Delta_{i_1 i_2 i_3 i_4}^{(1)}$ are independent of each other. Let $(j, s, m, \ell)$ be any cycle on the four nodes $\{i_1, i_2, i_3, i_4\}$. Then, the variance of $W_{js} \Omega_{sm} \Omega_{m\ell} \Omega_{\ell j}$ is bounded by $\Omega_{js} \Omega_{sm}^2 \Omega_{m\ell}^2 \Omega_{\ell j}^2 = O(\theta_j^3 \theta_s^3 \theta_m^4 \theta_\ell^4)$. Hence,

$$\sum_{CC(I_n)} \mathrm{Var}(\Delta_{i_1 i_2 i_3 i_4}^{(1)}) \leq C \sum_{j,s,m,\ell} \theta_j^3 \theta_s^3 \theta_m^4 \theta_\ell^4$$
$$\leq C\|\theta\|_3^6 \|\theta\|_4^8 = o(\|\theta\|_3^6 \|\theta\|^8),$$

where the last inequality is from the condition (7) and the fact that $\|\theta\|_4^4 = (\sum_i \theta_i^4) \leq \theta_{\max}^2 (\sum_i \theta_i^2) = O(\|\theta\|^2) = o(\|\theta\|^4)$. We then look at the covariance between $\Delta_{i_1 i_2 i_3 i_4}^{(1)}$ and $\Delta_{i_1' i_2' i_3' i_4'}^{(1)}$. Let $(j, s, m, \ell)$ be any cycle on the four nodes $\{i_1, i_2, i_3, i_4\}$, and let $(j', s', m', \ell')$ be any cycle on the four nodes $\{i_1', i_2', i_3', i_4'\}$. As long as $\{j, s\} \neq \{j', s'\}$, the two terms $W_{js} \Omega_{sm} \Omega_{m\ell} \Omega_{\ell j}$ and $W_{j's'} \Omega_{s'm'} \Omega_{m'\ell'} \Omega_{\ell' j'}$ are independent, hence, their covariance is zero. If $\{j, s\} = \{j', s'\}$, their covariance is bounded by $\Omega_{js} \cdot$

$\Omega_{sm}\Omega_{m\ell}\Omega_{\ell j}\Omega_{s'm'}\Omega_{m'\ell'}\Omega_{\ell' j} = O(\theta_j^3\theta_s^3\theta_m^2\theta_\ell^2\theta_{m'}^2\theta_{\ell'}^2)$. As a result,

$$\sum_{CC(I_n)\times CC(I_n)} \text{Cov}(\Delta^{(1)}_{i_1i_2i_3i_4}, \Delta^{(1)}_{i'_1i'_2i'_3i'_4})$$
$$\leq C\sum_{j,s,m,\ell,m',\ell'}\theta_j^3\theta_s^3\theta_m^2\theta_\ell^2\theta_{m'}^2\theta_{\ell'}^2 \leq C\|\theta\|_3^6\|\theta\|^8.$$

Note that $G(\Omega) \asymp n^4 C_4 \asymp \|\theta\|^8$ by Lemma B.1. Additionally, from the condition (7), $\|\theta\|_3 = o(1)$. Hence, the above imply

$$\text{Var}\left(\sum_{CC(I_n)}\Delta^{(1)}_{i_1i_2i_3i_4}\right) \ll G(\Omega). \quad (69)$$

Combining (68)-(69) gives

$$\frac{1}{\sqrt{G(\Omega)}}\sum_{CC(I_n)}\Delta^{(1)}_{i_1i_2i_3i_4} \xrightarrow{p} 0. \quad (70)$$

We can consider other terms similarly. By direct calculations,

$$\text{Var}\left(\sum_{CC(I_n)}\Delta^{(2)}_{i_1i_2i_3i_4}\right) \leq \sum_{j,s,m,\ell}\Omega_{js}\Omega_{sm}\Omega_{m\ell}^2\Omega_{\ell j}^2$$
$$+ \sum_{\substack{j,s,m\\ \ell,\ell'}}\Omega_{js}\Omega_{sm}\Omega_{m\ell}\Omega_{m,\ell'}\Omega_{\ell j}\Omega_{\ell' j}$$
$$\leq C\sum_{j,s,m,\ell}\theta_j^3\theta_s^2\theta_m^3\theta_\ell^4 + C\sum_{\substack{j,s,m\\ \ell,\ell'}}\theta_j^3\theta_s^2\theta_m^3\theta_\ell^2\theta_{\ell'}^2$$
$$\leq C\|\theta\|_3^6\|\theta\|^2\|\theta\|_4^4 + C\|\theta\|_3^6\|\theta\|^6 = o(\|\theta\|^8),$$

and

$$\text{Var}\left(\sum_{CC(I_n)}\Delta^{(3)}_{i_1i_2i_3i_4}\right) \leq \sum_{j,s,m,\ell}\Omega_{js}^2\Omega_{sm}\Omega_{m\ell}\Omega_{\ell j}$$
$$\leq C\sum_{j,s,m,\ell}\theta_j^3\theta_s^3\theta_m^2\theta_\ell^2$$
$$\leq C\|\theta\|_3^6\|\theta\|^4 = o(\|\theta\|^8).$$

Hence, for the terms related to $\Delta^{(2)}_{i_1i_2i_3i_4}$ and $\Delta^{(3)}_{i_1i_2i_3i_4}$, we also have a similar convergence as that of (70). These together imply $J_1 \xrightarrow{p} 0$. Hence, (66) is true.

Next, we prove (67). It is seen that

$$0 \leq G(\Omega) - G(\Omega^*) \leq C\sum_{j,s,m,\ell}\Omega_{js}^2\Omega_{sm}\Omega_{m\ell}\Omega_{mj}$$
$$\leq C\sum_{j,s,m,\ell}\theta_j^3\theta_s^3\theta_m^2\theta_\ell^2 \leq C\|\theta\|_3^6\|\theta\|^4 = o(\|\theta\|^8).$$

As a result, $|G(\Omega^*)/G(\Omega) - 1| = o(1)$. This proves (67). □

### D.4. Proof of Proposition A.1

The last item follows once the first three items are proved, so we only consider the first three items.

Consider the first item. Write

$$1'A^21 = \sum_{1\leq i_1,i_2,i_3\leq n}A_{i_1i_2}A_{i_2i_3}.$$

Recall that all diagonal entries of $A$ are 0, we can exclude the case $i_1 = i_2$ or $i_2 = i_3$ from the summation. Therefore, we only need to sum over either the cases where $i_1, i_2, i_3$ are distinct and the cases $i_1 = i_3$ but $i_1 \neq i_2$. It follows

$$1'A^21 = \left(\sum_{\substack{i_1,i_2,i_3\\ \text{are distinct}}} + \sum_{\substack{i_1,i_2,i_3\\ i_1=i_3, i_1\neq i_2}}\right)A_{i_1i_2}A_{i_2i_3} = I + II.$$

Now, first, by definition,

$$I = B_{n,3}\widehat{L}_2, \qquad \text{where } B_{n,3} = 6\binom{n}{3},$$

and second (recall all diagonal entries of $A$ are 0),

$$II = \sum_{i_1,i_2}A_{i_1i_2}^2 = \text{tr}(A^2).$$

Combining these gives

$$\widehat{L}_2 = \frac{1}{6\binom{n}{3}}(1'A^21 - \text{tr}(A^2)),$$

and the claim follows.

Consider the second item. Using similar arguments, we decompose

$$1'A^31 = \sum_{1\leq i_1,i_2,i_3,i_4\leq n}A_{i_1i_2}A_{i_2i_3}A_{i_3i_4} = I+II+III+IV,$$

where

$$I = \sum_{\substack{i_1,i_2,i_3,i_4\\ \text{are distinct}}}A_{i_1i_2}A_{i_2i_3}A_{i_3i_4} = B_{n,4}\widehat{L}_3,$$

with $B_{n,4} = 24\binom{n}{4}$,

$$II = \left(\sum_{\substack{i_1,i_2,i_3,i_4\\ i_1=i_3}} + \sum_{\substack{i_1,i_2,i_3,i_4\\ i_2=i_4}}\right)A_{i_1i_2}A_{i_2i_3}A_{i_3i_4} = 2\cdot(1'A^21),$$

$$III = \sum_{\substack{i_1,i_2,i_3,i_4\\ i_1=i_4}}A_{i_1i_2}A_{i_2i_3}A_{i_3i_4} = 1'A^31,$$

and

$$IV = -\sum_{\substack{i_1,i_2,i_3,i_4\\ i_1=i_3, i_2=i_4}}A_{i_1i_2}A_{i_2i_3}A_{i_3i_4} = -1'A1.$$

Combining these gives

$$\widehat{L}_3 = \frac{1}{24\binom{n}{4}}[1'A^3 1 - 2 \cdot 1'A^2 1 + 1'A1 - \text{tr}(A^3)],$$

and the claim follows.

Consider the third item. Note first

$$\text{tr}(A^4) = \sum_{1 \le i_1, i_2, i_3, i_4 \le n} A_{i_1 i_2} A_{i_2 i_3} A_{i_3 i_4} A_{i_4 i_1}.$$

Similarly, we have

$$\text{tr}(A^4) = \sum_{i_1, i_2, i_3, i_4} A_{i_1 i_2} A_{i_2 i_3} A_{i_3 i_4} A_{i_4 i_1} = I + II + III,$$

where

$$I = \sum_{\substack{i_1, i_2, i_3, i_4 \\ \text{are distinct}}} A_{i_1 i_2} A_{i_2 i_3} A_{i_3 i_4} A_{i_4 i_1} = 24 \binom{n}{4} \widehat{C}_4,$$

$$II = \left( \sum_{\substack{i_1, i_2, i_3, i_4 \\ i_1 = i_3}} + \sum_{\substack{i_1, i_2, i_3, i_4 \\ i_2 = i_4}} \right) A_{i_1 i_2} A_{i_2 i_3} A_{i_3 i_4} A_{i_4 i_1} = 2 \cdot (1'A^2 1)$$

and

$$III = - \sum_{\substack{i_1, i_2, i_3, i_4 \\ i_1 = i_3, i_2 = i_4}} A_{i_1 i_2} A_{i_2 i_3} A_{i_3 i_4} A_{i_4 i_1} = -1'A1.$$

Combining these gives

$$\widehat{C}_4 = \frac{1}{24\binom{n}{4}} \left( tr(A^4) - 2 \cdot 1'A^2 1 + 1'A1 \right),$$

and the claim follows. □